\documentclass[fleqn,usenatbib]{mnras}

\usepackage{newtxtext,newtxmath}

\usepackage[T1]{fontenc}
\usepackage{ae,aecompl}
\usepackage{mathtools}

\usepackage{graphicx}	
\usepackage{amsmath}	
\usepackage{siunitx} 
\usepackage{makecell}

\usepackage{CJKutf8}

\newcommand{\msun}{$M_{\mathrm{\odot}}$}

\newdimen\digitwidth    
\setbox1=\hbox{0}       
\digitwidth=\wd1        
\catcode`"=\active      
\def"{\kern\digitwidth}

\def\kms{km~s$^{-1}$}
\def\msun{$M_{\odot}$}
\def\HI{H\,{\textsc{\romannumeral 1}}}
\def\HII{H\,{\textsc{\romannumeral 2}}}
\def\NII{N\,{\textsc{\romannumeral 2}}}
\def\Ha{H$\alpha$}

\newcommand{\angstrom}{\text{\normalfont\AA}}

\title{On the origin of the anomalous gas, non-declining rotation curve and disc asymmetries in NGC 253}

\author[Xuanyi Lyu et al.]{
Xuanyi Lyu 
\begin{CJK*}{UTF8}{gbsn}
(吕宣懿)
\end{CJK*}$^{1}$\thanks{E-mail: xuanyi.lyu.astro@gmail.com},
T. Westmeier$^{1}$\thanks{E-mail: tobias.westmeier@uwa.edu.au},
Gerhardt R. Meurer$^{1}$, D. J. Hanish$^{2}$
\\
$^{1}$International Centre for Radio Astronomy Research, The University of Western Australia, 35 Stirling Highway, Crawley, WA 6009, Australia\\
$^{2}$Spitzer Science Center, California Institute of Technology, MC 220-6, 1200 E California Blvd., Pasadena, CA 91125, USA\\
}

\date{Accepted 2023 June 9. Received 2023 June 2; in original form 2023 February 28}

\pubyear{2015}

\begin{document}
\label{firstpage}
\pagerange{\pageref{firstpage}--\pageref{lastpage}}
\maketitle

\begin{abstract}
We present a multi-wavelength (from far ultraviolet to \HI\ emission) study of star formation feedback on the kinematics of the interstellar medium in the Sculptor Galaxy, NGC~253. Its three well-known features (a disrupted stellar disc, a previously reported delining rotation curve, and anomalous \HI\ gas) are studied in a common context of disc asymmetries. About 170 h of on-source ATCA observations are collected and reduced into two versions of \HI\ data cubes of different angular resolution (30$\arcsec$ / 2$\arcmin$) and HI column density sensitivity (7.4$\times$ $10^{19}$~cm$^{-2}$ / 4$\times$ $10^{18}$~cm$^{-2}$). We separate the anomalous gas from the disc using a custom-made line profile fitting toolkit called FMG. Two star formation tracers (H$\alpha$, FUV emission) are carefully processed and studied. We find that at $R > 7.5~\mathrm{kpc}$ the star formation activity is strongly lopsided (SFR$_{NE}$ >SFR$_{SW}$), and investigate several other properties (H$\alpha$/FUV, dust temperature, stellar age, and disc stability parameters). We also find that the declining nature of the rotation curve perceived by previous studies is not intrinsic but a combined effect of kinematical asymmetries at $R = 7.5$--$16~\mathrm{kpc}$. This is likely the consequence of star formation triggered outflow. The mass distribution and the timescale of the anomalous gas also imply that it originates from gas outflow, which is perhaps caused by galaxy-galaxy interaction considering the crowded environment of NGC~253.

\end{abstract}
\begin{keywords}
galaxies: individual: NGC~253 -- galaxies: ISM -- radio lines: galaxies -- galaxies: kinematics and dynamics -- galaxies: star formation -- galaxies: starburst
\end{keywords}



\section{Introduction}

Stellar feedback, especially feedback from massive stars, plays a key role in the evolution of the interstellar medium (ISM). Stellar winds and supernovae expelled energy and matter from the disc to the halo, which changes the structure and abundance of the gaseous halo. This process has been observed in a sample of nearby galaxies at multiple wavelengths (eg. NGC 891: \cite{Oosterloo2007, Mouhcine2010}, M33: \cite{Putman2009, Grossi2008}). NGC~253 (the basic information is summarized in Table~\ref{tab:basic_info}) is one of the most active star-forming systems in the local universe. It is suggested that the inner $\sim$ 300pc region has a star formation rate of $\sim$ 3.5\msun\ yr$^{-1}$ and a supernovae rate of $\sim$ 0.2 yr$^{-1}$ \citep[]{Bendo2015_1, Rampadarath2014}. The burst-driven outflow, which originates from the star-forming core, has been extensively studied (X-rays: \cite{Strickland2000, Strickland2002, Bauer2008}, \Ha:\ \cite{Westmoquette2011}, and molecular gas: \cite{Turner1985, Bolatto2013, Walter2017, Krieger2019, Levy2021}). The outflow rate is estimated to be comparable with the SFR of the inner region, which in turn limits the star formation activities in the centre.

\begin{table}
	\centering
	\caption{Basic information of NGC 253}
	\label{tab:basic_info}
	\begin{tabular}{lccr} 
		\hline
		  Parameter  & Value\\
		\hline
		Right ascension (J2000) & $00^{\si{h}} 47^{\si{m}} 33.1^{\si{s}}$\\
		Declination (J2000) & $\ang{-25;17;17.9}$\\
		Morphological type & SABc\\
		Distance & 3.94 Mpc\\
        Systemic velocity & 238 $\pm{~4}$\kms\ \\
		D$_{25}$ &  $31.5~\mathrm{kpc}$\\
		Inclination & $\ang{78.5}$\\
		Position angle & $\ang{52}$\\
		log M$_{stars}$ (M$_\odot$) & 10.33\\
		log M$_{\HI\ }$ (M$_\odot$) & 9.44\\
        SFR of the whole galaxy & 3.86 \(M_\odot\)\si[per-mode=symbol]{yr^{-1}} \\
        \makecell[l]{SFR of outer disk \\ (radius $\geqslant$ 2.5$~\mathrm{kpc}$)} & 1.84 \(M_\odot\)\si[per-mode=symbol]{yr^{-1}}\\
        
		\hline
	\end{tabular}
\end{table}

In addition to the inner bar and core, previous studies have shown that NGC~253 has another three interesting features. 

\subsection{Previously reported declining rotation curve}
The rotation curve (RC) is a key tool to understand the mass distribution of galaxies. Most spiral galaxies have a flat or rising rotation curve across their outer disc, which results from the contribution of massive dark matter halos surrounding the discs. However, a few exceptions were found in the past decades (NGC~5907: \cite{Casertano1983}; NGC~7793: \cite{Carignan1990}, \cite{Dicaire2008}; NGC~2683 \& NGC~3521: \cite{Casertano1991}). Their rotation velocity decreases by $\sim$ 25 \% of the maximum velocity at large radii. NGC~253 is one of those exceptions. The declining trend of rotation velocity has previously been reported in both the neutral (\HI)\ and ionised (\Ha)\ gas disc at a radius of $R>12\arcmin$. The \HI\ RC of NGC~253 was first measured by \cite{Puche1991} using VLA observations. Due to limited sensitivity, they only derive the RC up to a radius of $\sim$12\arcmin\ ($\sim$14$~\mathrm{kpc}$), and find that the RC keeps rising until the last observed velocity point. Subsequently, by using TAURUS-2 Fabry-P\'{e}rot interferometer observation of \Ha\ and [\NII\ ] emission lines in the outer parts of NGC~253 (in the southeast side) \cite{Bland-Hawthorn1997} report a velocity decrease of $\sim$10 percent in the outskirts compared to $V_{max}$ measured by \cite{Puche1991}. \cite{Hlavacek-Larrondo2011} present a deep \Ha\ kinematical analysis by using the same observation technique as \cite{Bland-Hawthorn1997} and report that the RC drops by $\sim$ 30 percent compared to $V_{max}$ at R$\sim$ 11 \arcmin.\ Recently, \cite{Lucero2015} observed \HI\ emission of NGC~253 using the MeerKAT array (Karoo Array Telescope, KAT-7). They measure the RC out to a radius of $\sim$ 18 \arcmin\ and reproduce the declining trend reported by \cite{Bland-Hawthorn1997} and \cite{Hlavacek-Larrondo2011} in \Ha\ emission.

A declining RC provides vital insights into galaxy evolution since it not only breaks the conspiracy between the baryonic component and dark matter to maintain a flat RC but also implies a truncated dark matter halo close to the \HI\ edge. Thus, we try to obtain a more detailed analysis of the \HI\ RC using higher angular resolution and better sensitivity ($\sim$ 30\arcsec;\ 0.88 mJy/beam) than \cite{Puche1991} ($\sim$ 1\arcmin;\ 6.9 mJy/beam).

\subsection{Anomalous gas}
As introduced before, extra-planar \HI\ has been detected in a few nearby galaxies. NGC~253 is one of them. \cite{Boomsma2005} first found a diffuse extraplanar \HI\ structure extending up to 12$~\mathrm{kpc}$ vertically from the disc, which is confirmed by \cite{Lucero2015}. Both studies separate anomalous gas from the disc by visual inspection and give a similar estimation of the total mass of the anomalous \HI\ ($\sim$ 8 $\times$ $10^{7}$ \msun\ ), which is around 3 percent of the total \HI\ mass. Although both \cite{Boomsma2005} and \cite{Lucero2015} prefer that the anomalous \HI\ is outflowing gas, the origin of the anomalous \HI\ remains uncertain. As a result, we try to study the anomalous gas using a better method (Gaussian decomposition analysis of the \HI\ profile of each pixel) and more sensitive observation (3 $\sigma\ $ \HI\ column density threshold of $\sim$4$\times$ $10^{18}$~cm$^{-2}$).
    
\subsection{Extended and probably disrupted stellar disc}

Most spiral galaxies have much more extended \HI\ envelopes than their stellar components. However, the size of the \HI\ disc ($\sim$ 0.8$\times$ Holmberg radius) in NGC~253 is similar to the optical one. This is neither because NGC~253 is an outlier on the \HI\ mass-size relation \citep{Broeils1997}, nor a consequence of \HI\ deficiency \citep{Lucero2015}. The only reason left is that the stellar disc of NGC~253 is unusually extended. \cite{Davidge2010} found that the disc is traced by the bright asymptotic giant branch (AGB) stars out to at least 13 scale lengths (22$~\mathrm{kpc}$) along the major axis. In addition, there is an extended extraplanar stellar component, which possibly resulted from tidal interactions with a companion. Recently in NGC~247, one of the companions of NGC~253, structures such as voids and bubbles were found in UV and near-infrared images after deprojection \citep{Davidge2021}, implying a recent interaction with NGC~253. Moreover, elevated levels of star formation rate in the disc, as an evident consequence of galaxy-galaxy interaction, were also detected by \cite{Davidge2010}. The highly populated red supergiants (RSGs) in the northern parts of NGC~253 indicate that local intensive star formation activities occurred recently ($\sim$ several 10 Myr ago), which spatially coincides with the anomalous \HI\ found by \cite{Boomsma2005}. 

Inspired by this idea, we study the spatial distribution of star formation activity in NGC~253's disc using FUV and \Ha\ observations, and the ratio between them. FUV and H$\alpha$ are both commonly used star formation tracers. They reflect different mass ranges of stars: FUV traces both O and B stars (M$_{\ast}$ $\geq$ 3\(M_\odot\)), while H$\alpha$ is only emitted by O stars (M$_{\ast}$ $\geq$ 20\(M_\odot\)). Many former studies take advantage of this property to study the initial mass function (IMF) or star formation history (SFH). The difference in FUV to H$\alpha$ ratio is either caused by a nonuniversal IMF (the number ratio of two mass ranges of stars is not fixed) or caused by different recent SFH (B stars have $\sim$100Myr lifetime while O stars last less than 5Myr) or a combination of both. In this study, we will calculate the unattenuated FUV to H$\alpha$ flux ratio and its correlation with ISM properties. However, we will not explore the possible explanations for the variation of this ratio, which is so complex that it requires direct constraints on the local SFH.

\subsection{Disc asymmetries}

Of particular interest is that all three features mentioned above are tightly linked to the same fact: the disc asymmetries of NGC~253. Firstly, all previously reported declining RCs of NGC~253, no matter whether it is derived from \HI\ or \Ha\ observations, suggest an asymmetric disc. The differences in rotation velocity between the approaching side (NE) and the receding side (SW) of the disc are significant in the outer disc (shown in Figure 6 in \cite{Hlavacek-Larrondo2011}, Figure 13 in \cite{Lucero2015}; a similar trend could also be observed in Figure 8 of \cite{Puche1991} although a rising RC is concluded there due to limited sensitivity of their observation). \cite{Lucero2015} mentioned that much more \HI\ signal is detected on the receding side. \cite{Hlavacek-Larrondo2011} also suggests that there is no extended diffuse ionized gas on the approaching side, which is found on the receding side. Second, both \cite{Boomsma2005} and \cite{Lucero2015} indicate that the distribution of the anomalous gas is strongly asymmetric. Most anomalous gas is found on the NE side of the disc and is bisected by the northeast side of the disc, where the star formation is boosted (Davidge 2010). Many other spatial distributions studied by \cite{Davidge2010} are also lopsided. For example, the number density of M giants is higher in the NE quadrant, as is the density of AGB stars. 

We try to solve several fundamental issues highlighted by these features: Is the star formation more active in the northeast quadrant than on the southwest side? Why is the rotation curve perceived to be declining in the outskirts of the disc? What is the origin of anomalous gas? And most importantly, what is the reason for the asymmetry of these features? To answer these questions, we gathered multi-wavelength observations (from far ultraviolet to 21-cm radio emission) to study the spatial correlation between \HI\ kinematics (rotation curve), anomalous gas, and star formation activities in the context of disc asymmetries. 

This paper is organized as follows: Section~\ref{section:data} describes the data reduction and processing for \HI,\ H$\alpha$, FUV, and infrared observations. In Section~\ref{section:HI} we discuss the rotation curve fitting process. Also, the Gaussian decomposition method we developed and the results of applying it to our \HI\ data will be described in this section. The main results will be discussed in Section~\ref{section:results} and summarized in Section~\ref{section:conclusions}.

\section{Data}
\label{section:data}

\begin{table}
	\centering
	\caption{Basic information of multi-wavelength data}
	\label{tab:multiband_obs}
	\begin{tabular}{lccr} 
		\hline
		  wavelength  & instruments & FWHM\\
		\hline
		FUV(1524$\angstrom$) & GALEX & 4.48$\arcsec$ \\
		NUV(2297$\angstrom$) & GALEX & 4.48$\arcsec$\\
		H$\alpha$(6563$\angstrom$) & CTIO & 1.97$\arcsec$\\
		G band(330-1050nm) & Gaia & -$^{(a)}$\\
		$\SI{3.4}{\micro\metre}$ & WISE(w1) & 5.79$\arcsec$\\
		$\SI{70}{\micro\metre}$ & Herschel(PACS) & 5.67$\arcsec$\\
		$\SI{100}{\micro\metre}$ & Herschel(PACS) & 7.04$\arcsec$\\
		$\SI{160}{\micro\metre}$ & Herschel(PACS) & 11.18$\arcsec$\\
		\HI\ (21cm) & ATCA & $\sim$0.5$\arcmin$/2$\arcmin$\\

		\hline
	\end{tabular}
	

Notes. $^{(a)}$ for Gaia data, only G band magnitude and astrometric information were used. 
      
\end{table}

\begin{figure*}
	\includegraphics[width=1.\textwidth]{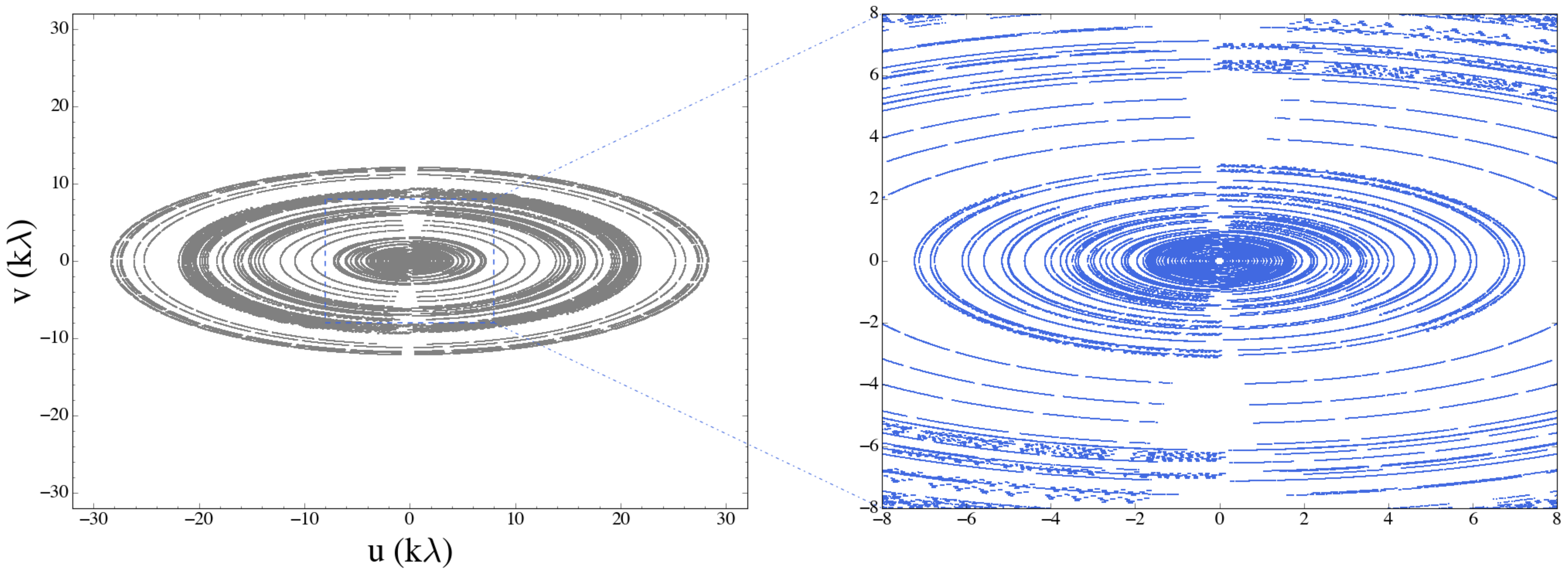}
    \caption{uv-coverage of all 17 epochs. the right panel is a zoom-in view of the centre part.}
    \label{fig:fig_uv}
\end{figure*}

\subsection{\HI\ data reduction}

We collected all \HI\ observations of NGC~253 taken by the Australia Telescope Compact Array (ATCA) between 1993 and 2020 from the Australia Telescope Online Archive\footnote{\url{https://atoa.atnf.csiro.au/}}. After detailed checking, 17 epochs of observations were finally adopted, the details of which are given in Table~\ref{tab:hi_obs}. For most of the epochs, all six 22-meter antennas were kept for better u-v coverage. The final u-v coverage is shown in Fig.~\ref{fig:fig_uv}. The total on-source integration time is 168 h. In addition to the centre of NGC~253, multiple pointings surrounding the major axis of the disc were also retrieved. This increases the sensitivity to extended structures considering the limited angular size of the ATCA primary beam compared to the size of NGC~253. As a result, a nearly flat sensitivity distribution is obtained for the area of interest. 

\begin{table*}
    	\centering
	    \caption{Details of ATCA observations}
    	\label{tab:hi_obs}
    	
	    \begin{tabular}{ccccccc} 
		    \hline
		 Observing date & Project code & \makecell[c]{Velocity resolution \\( \kms\ )} & Array configurations & correlator & pointings & \makecell[c]{On-source time \\(h)}	\\
		    \hline
         \date{mydate}{2013-}{12-}{07} & C2771 & 0.1  & 750B  & CABB & 6 & 6.8 \\
         \date{mydate}{2013-}{11-}{30} & C2771 & 0.1  & EW352 & CABB & 6 & 7.5 \\
         \date{mydate}{2013-}{11-}{17} & C2771 & 6.59 & EW352 & CABB & 6 & 10.9 \\
         \date{mydate}{2013-}{11-}{16} & C2771 & 6.59 & EW352 & CABB & 6 & 9.1 \\
         \date{mydate}{2013-}{11-}{15} & C2771 & 6.59 & EW352 & CABB & 6 & 10.6 \\
         \date{mydate}{2012-}{12-}{02} & C2771 & 6.59 & 1.5C  & CABB & 6 & 9.1 \\
         \date{mydate}{2012-}{12-}{01} & C2771 & 6.59 & 1.5C  & CABB & 6 & 9.2 \\
         \date{mydate}{2012-}{11-}{30} & C2771 & 6.59 & 1.5C  & CABB & 6 & 9.4 \\
         \date{mydate}{2006-}{07-}{24} & CX117 & 3.29 & H168  & old  & 5 & 7.5 \\
         \date{mydate}{2005-}{10-}{06} & C1341 & 3.29 & EW214 & old  & 16 & 10  \\
         \date{mydate}{2002-}{09-}{30} & C1025 & 3.29 & EW367 & old  & 1 & 11.5  \\
         \date{mydate}{2002-}{08-}{06} & C1025 & 3.29 & 750B  & old  & 1 & 9.7  \\
         \date{mydate}{2002-}{07-}{10} & C1025 & 3.29 & 1.5G  & old  & 1 & 11.3  \\
         \date{mydate}{1995-}{08-}{03} & C296  & 3.29 & 375   & old  & 1 & 11.8  \\
         \date{mydate}{1994-}{02-}{28} & C296  & 3.29 & 750A  & old  & 1 & 10.4  \\
         \date{mydate}{1994-}{01-}{15} & C296  & 3.29 & 6A    & old  & 1 & 11.3  \\
         \date{mydate}{1993-}{08-}{26} & C296  & 3.29 & 1.5B  & old  & 1 & 11.6  \\

		    \hline
	    \end{tabular}

\end{table*}

For each epoch of observation, the data reduction was conducted separately using a combination of the MIRIAD software package \citep{Sault1995} and Common Astronomy Software Applications \citep[CASA]{McMullin2007}. Flagging, bandpass calibration, and gain calibration were done using MIRIAD. The radio frequency interference (RFI) was carefully checked and removed for observations with compact array configurations (especially those with baselines shorter than 100 metres). It is important to note that there is some possible contamination from  Galactic \HI\ emission, especially in the low-resolution data, where the column density sensitivity reaches $\sim$4 $\times$ $10^{18}$~cm$^{-2}$. 
We searched for the Galactic \HI\ in the data cube of the Parkes observation of the Sculptor group by \cite{Westmeier2017} since single-dish observations would not suffer from the short-spacing problem and therefore detect all emission, including Galactic gas. We find that the Galactic contamination at the position of NGC~253 is mainly located at a velocity range of $-40$ to $+50~$ \kms\ (NGC~253's velocity range: 0-500 \kms.)\ 
After checking at this velocity range, we did not find obvious Galactic contamination in the two versions of the \HI\ data cube. This is probably caused by the limited capability of interferometer observations to detect extended emissions on large angular scales. 

The data sets were then imported to CASA for self-calibration for two reasons: 
\begin{enumerate}
\item Since the continuum image of NGC~253 has an extremely high dynamic range (the central core has a signal-to-noise ratio $> 1000$) and contains components on different angular scales, the multi-scale cleaning algorithm is necessary, which is not available in MIRIAD.
\item The non-coplanar baseline effect introduced by long baselines, especially those above 10 $\mathrm{k}\lambda\ $, is no longer negligible. To properly describe the three-dimensional u-v-w coverage, the w-projection needs to be taken into account, which is also not supported in MIRIAD.
\end{enumerate}
The self-calibration started with phase calibration and was repeated several times with a decreasing cleaning threshold and gain interval. Both phases and amplitudes were calibrated in the final iteration, where the time interval was set to 1 minute. Subsequently, continuum emission was subtracted by fitting 2nd-order polynomials to the line-free channels. The resulting visibility data were imaged, deconvolved, and restored using the CASA task clean. 300 w-planes were calculated for w-projection imaging. A set of 3 scales (single pixel, FWHM, and 3 $\times $ FWHM ) were used for multiscale deconvolution. The ``smallscalebias'' parameter was set as 0.5. 

Finally, two versions of \HI\ data cubes were generated by using different imaging parameters (details in Table~\ref{tab:data_products}). At a Briggs weighting robustness of 0.25, our high-resolution data produces an angular resolution of $\sim$30\arcsec\ at a velocity resolution of 8~\kms.\ The low-resolution data cube was generated by using a Briggs weighting robustness of 0.75 and adding a 1 arcmin Gaussian tapering to enhance the sensitivity, resulting in a final beam size of $\sim$1.5$\times$2.5 \arcmin.\ The HI emission of NGC 253 was extracted with the Source Finding Application \citep[SoFiA; ][]{Serra2015, Westmeier2021} by applying the smooth+clip algorithm and a detection threshold of 5$\sigma$ for the high-resolution cube and 6$\sigma$ for the low-resolution cube. The velocity field (top panel of Fig.~\ref{fig:moments}) of the high-resolution data was also generated by SoFiA for rotation curve fitting, which is discussed in section~\ref{subsection:RC_fitting}.

\begin{table}
    	\centering
	    \caption{Two versions of data products}
    	\label{tab:data_products}

	    \begin{tabular}{ccc} 
		\hline
		    & high-resolution & low-resolution\\
		\hline
		    Briggs weighting robustness & 0.25 & 0.75 \\

		    FWHM of tapering  & none & $\sim$60\arcsec\  \\

		    Angular resolution & 36\arcsec$\times$21\arcsec\ & 154\arcsec$\times$81\arcsec\ \\

            RMS noise  & 0.88 mJy beam$^{-1}$ & 0.72 mJy beam$^{-1}$ \\

		    \makecell[c]{column density sensitivity \\ 3$\sigma$ in 20~\kms\ } & 7.4 $\times$ $10^{19}$~cm$^{-2}$  & 3.8 $\times$ $10^{18}$~cm$^{-2}$ \\
        \hline
	    \end{tabular}
	
\end{table}

\subsection{Data processing for FUV and H$\alpha$ observation}

Two star-formation tracers were studied. H$\alpha$ observations were obtained from the Survey of Ionization in Neutral Gas Galaxies \citep[SINGG; ]{Meurer2006}, and the FUV image was collected from the NASA/IPAC Extragalactic Database (NED). To properly remove foreground stars and correct the dust attenuation, multi-wavelength data was also used. The ultraviolet to infrared observations involved in this project are summarized in Table~\ref{tab:multiband_obs}.

\subsubsection{Foreground star subtraction}
\label{subsubsection:star_sub}
\begin{figure}
	\includegraphics[width=\columnwidth]{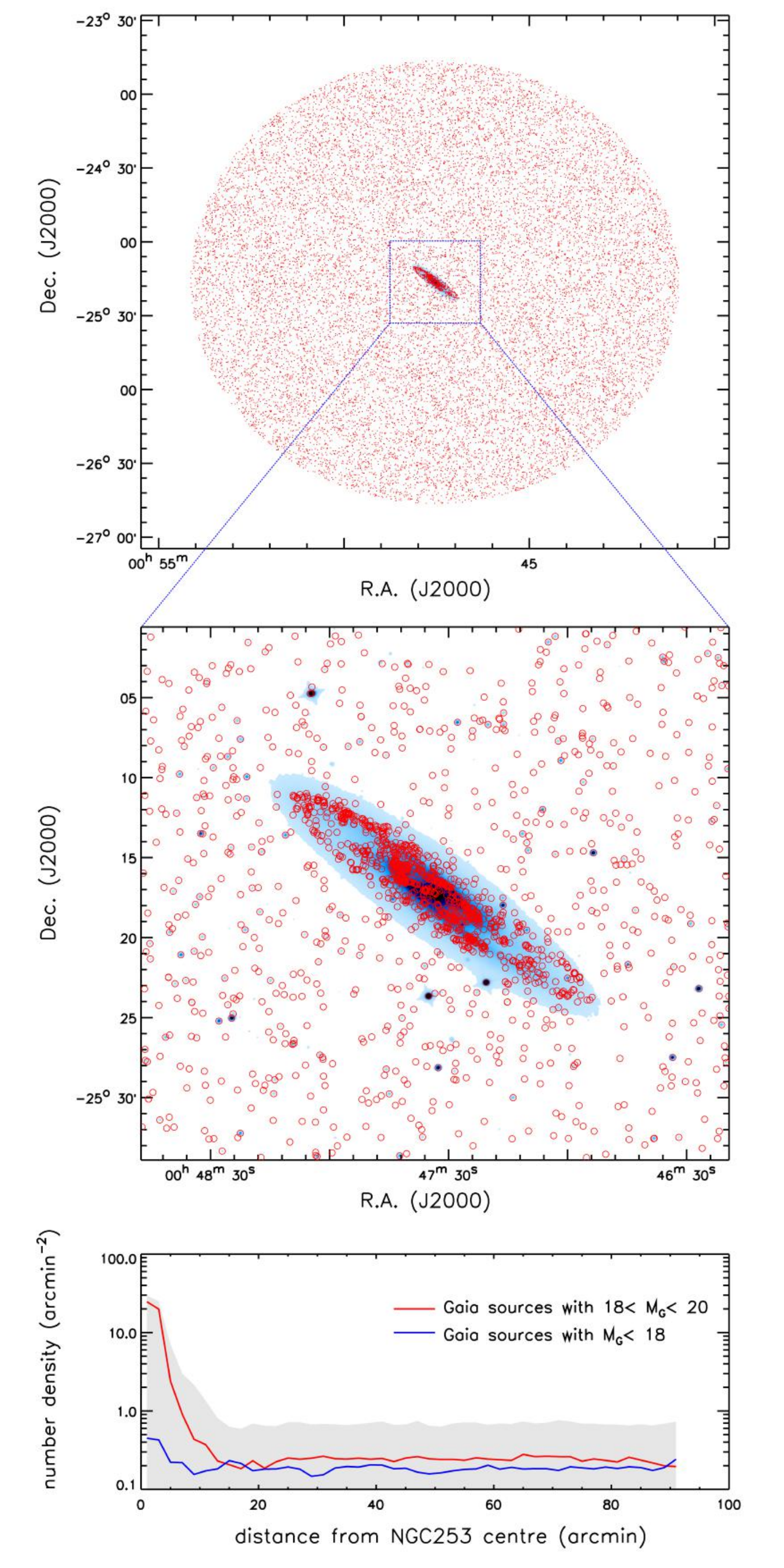}
    \caption{Top panel: the Gaia sources (in red dots) within 1.5$^{\circ}$ of NGC~253 plotted on a WISE band-1 image (blue).
    Middle panel: same as the top panel, but zoomed into NGC 253. 
    Bottom panel: the profile of Gaia source number density in arcmin$^{-2}$, which is estimated using the distance to the 5th nearest neighbour. The grey background shows how the density of all Gaia sources changes with distance to the centre of NGC~253. The blue and red lines respectively show the density of sources with M$_{G}$ < 18 and 18 < M$_{G}$ $\leqslant$ 20.}
    \label{fig:Gaia_sources}
\end{figure}

Gaia Data Release 2 \citep{Gaia2018} provides excellent astrometry and photometry measurements for point sources down to ~21 mag in the G band. However, a simple removal of all the Gaia sources does not work in this case. NGC~253 is a very nearby galaxy and many structures within its disc, such as star clusters, are also recognized as point sources in the GAIA DR2 catalogue. The top and middle panels of Fig.~\ref{fig:Gaia_sources} display the distribution of Gaia sources on top of the WISE-1 image of NGC 253. An overdensity is clearly observed in NGC 253's disc, which suggests that a majority of Gaia sources in this field are not foreground stars. The bottom panel of Fig.~\ref{fig:Gaia_sources} shows the number density profile of Gaia sources, which suggests that the point sources of NGC~253 have two properties.

\begin{enumerate}
\item They are mainly found within a radius of 20 \arcmin\ from the centre of the galaxy. Over this radius, the Gaia sources have a constant number density of $\sim$0.7 arcmin$^{-2}$. Adopting the assumption that the Galactic stars are uniformly distributed in the nearby region of NGC~253 (radius < 1.5$^{\circ}$), we could test our final foreground star removal by testing how well it matches the background Gaia source number density.
\item They mostly consist of sources with G band magnitude greater than 18 since the number density of m$_{G}$<18 (shown as the blue line in the bottom panel of Fig.~\ref{fig:Gaia_sources}) is fairly constant. As a result, Gaia sources with m$_{G}$<18 should be recognized as Galactic stars and removed from the FUV and H$\alpha$ images.
\end{enumerate}

\begin{figure}
	\includegraphics[width=\columnwidth]{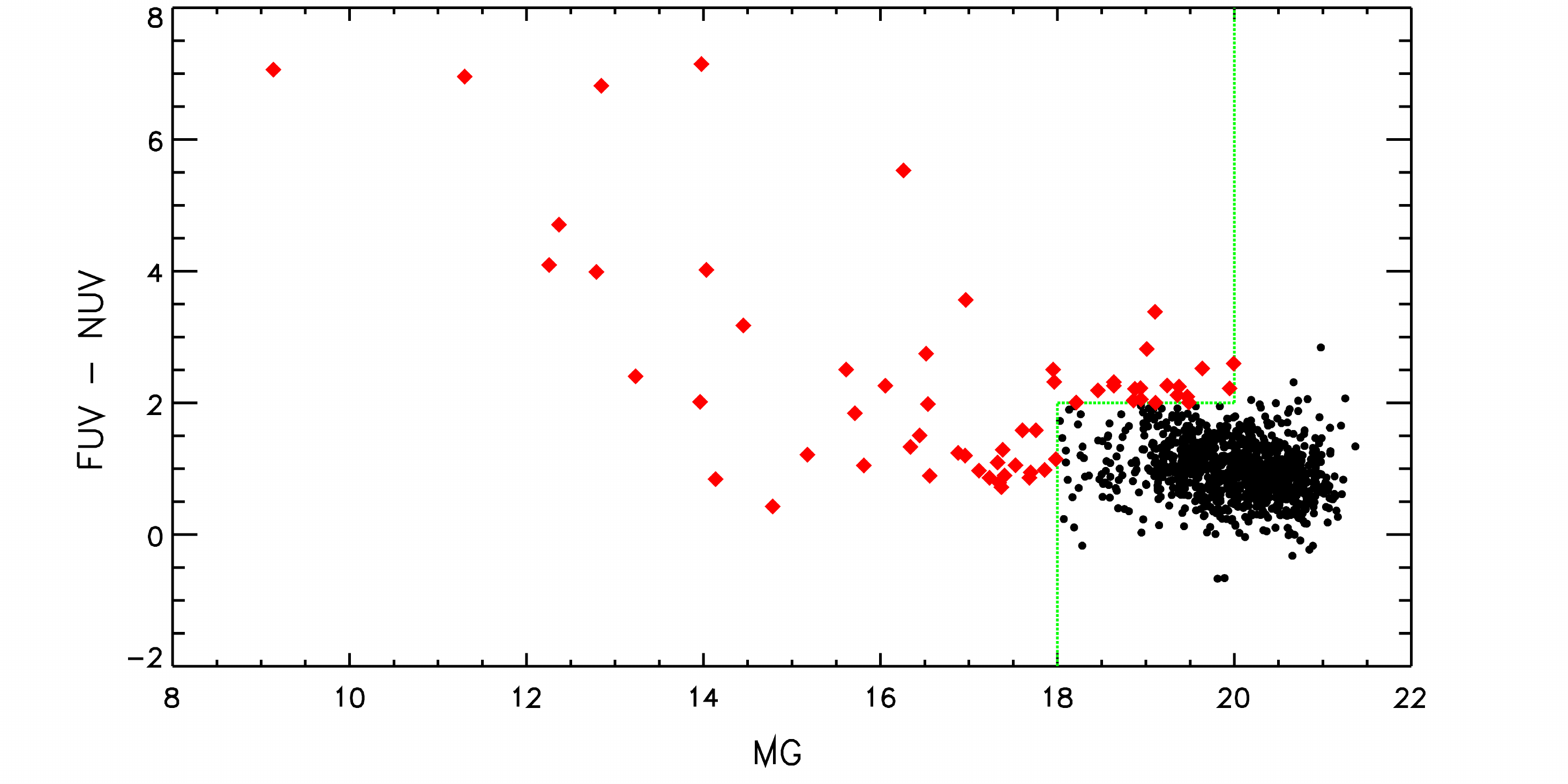}
    \caption{UV colour of Gaia sources within 20$~\mathrm{kpc}$ of NGC253 compared to their G band magnitude. The green line shows our selection criteria. Red dots are the Gaia sources recognized as foreground stars and black squares are the remaining Gaia sources.}
    \label{fig:star_removel_criteria}
\end{figure}

To remove the Galactic stars with 18 < m$_{G}$ $\leqslant$ 20, we adopt the method from \cite{Bianchi2007}, which successfully distinguished the remote objects from the Galaxy using the UV colour from GALEX. Figure 7 of \cite{Bianchi2007} shows that most non-Galactic objects have FUV-NUV colour smaller than 2. Following this, all Gaia sources within 20\arcmin\ of NGC~253 are divided into three subgroups by their G band magnitude:

\begin{enumerate}
\item m$_{G}$ $\leqslant$ 18mag: they are all recognized as foreground stars since no structures belonging to NGC253 should be this bright.
\item 18mag< m$_{G}$ $\leqslant$ 20mag: they are removed if their GALEX colour FUV-NUV > 2 following \cite{Bianchi2007}.
\item m$_{G}$> 20mag: we leave this subgroup alone. Although a small portion of them are still stars, they are too faint to influence the FUV/H$\alpha$ flux measurement.
\end{enumerate}

\begin{figure}
	\includegraphics[width=\columnwidth]{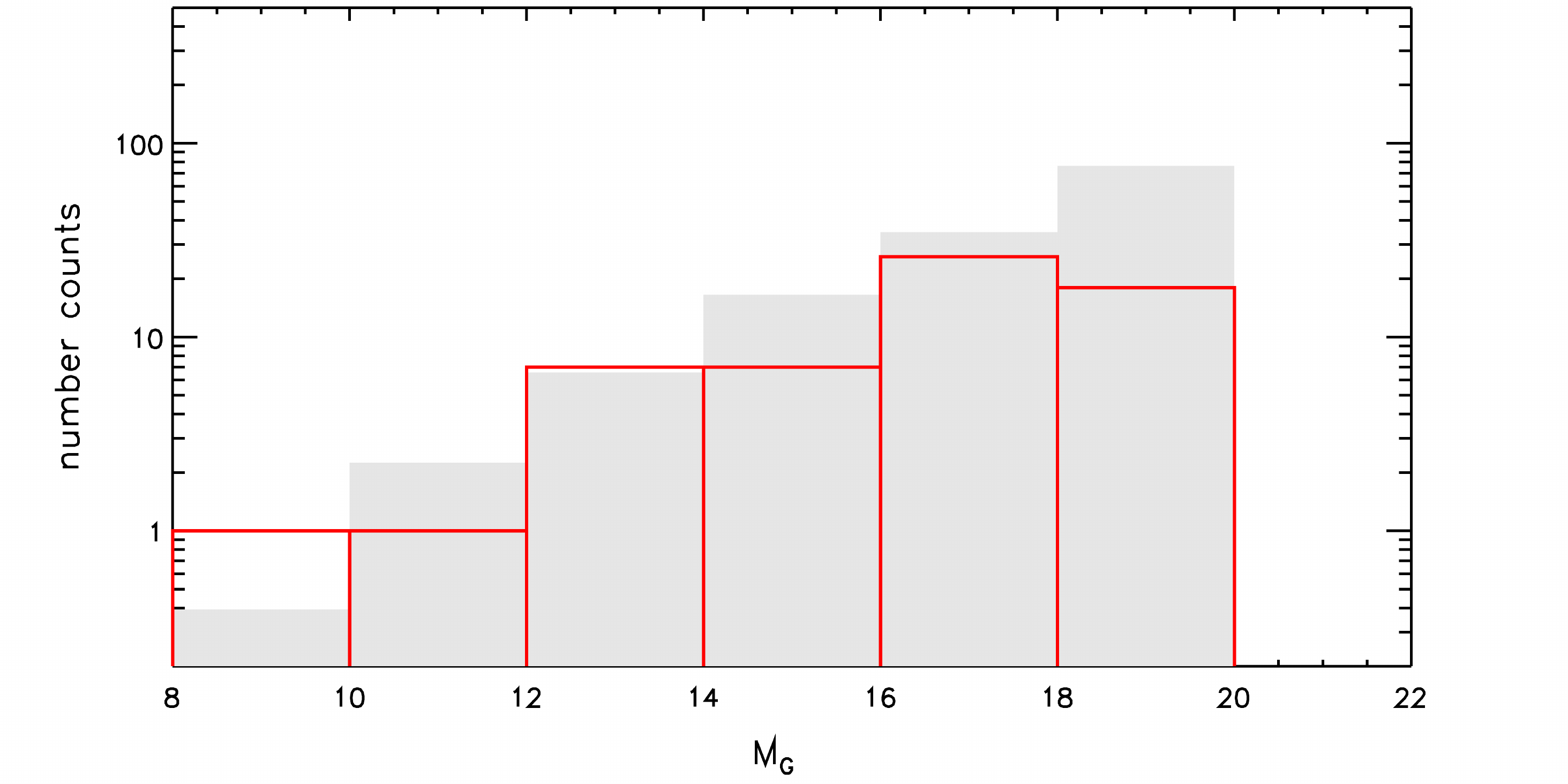}
    \caption{Histogram plot of the G magnitude distribution of removed stars. The grey background shows the number density of stars surrounding NGC~253 ( 1$^{\circ}$ < radius $\leqslant$ 1.4$^{\circ}$).}
    \label{fig:removed_stars}
\end{figure}

Following this algorithm, 60 objects deemed to be foreground stars have been removed. Fig.~\ref{fig:removed_stars} shows their magnitude distribution and the estimation from the nearby field, which suggests that most foreground stars with m$_{G}$ $\leqslant$ 19mag are recognized and removed by our criteria. Foreground star removal is performed using the task IMEDIT in IRAF.

\subsubsection{Dust attenuation and star formation}
\label{subsubsection:dust}
 
Dust attenuation for FUV and H$\alpha$ is corrected using the method developed by \cite{Kennicutt2009} and \cite{Hao2011}, which is a linear combination with total infrared luminosity (TIR). TIR was estimated using far infrared observation from the Herschel Space Observatory, following \cite{Galametz2013}. Usually, dust attenuation methods for star formation tracers in \cite{Kennicutt2009} and \cite{Hao2011} are developed for galaxies as a whole. Here, for simplicity, we have adapted this method for our spatially resolved study.

\begin{figure}
    \includegraphics[width=\columnwidth]{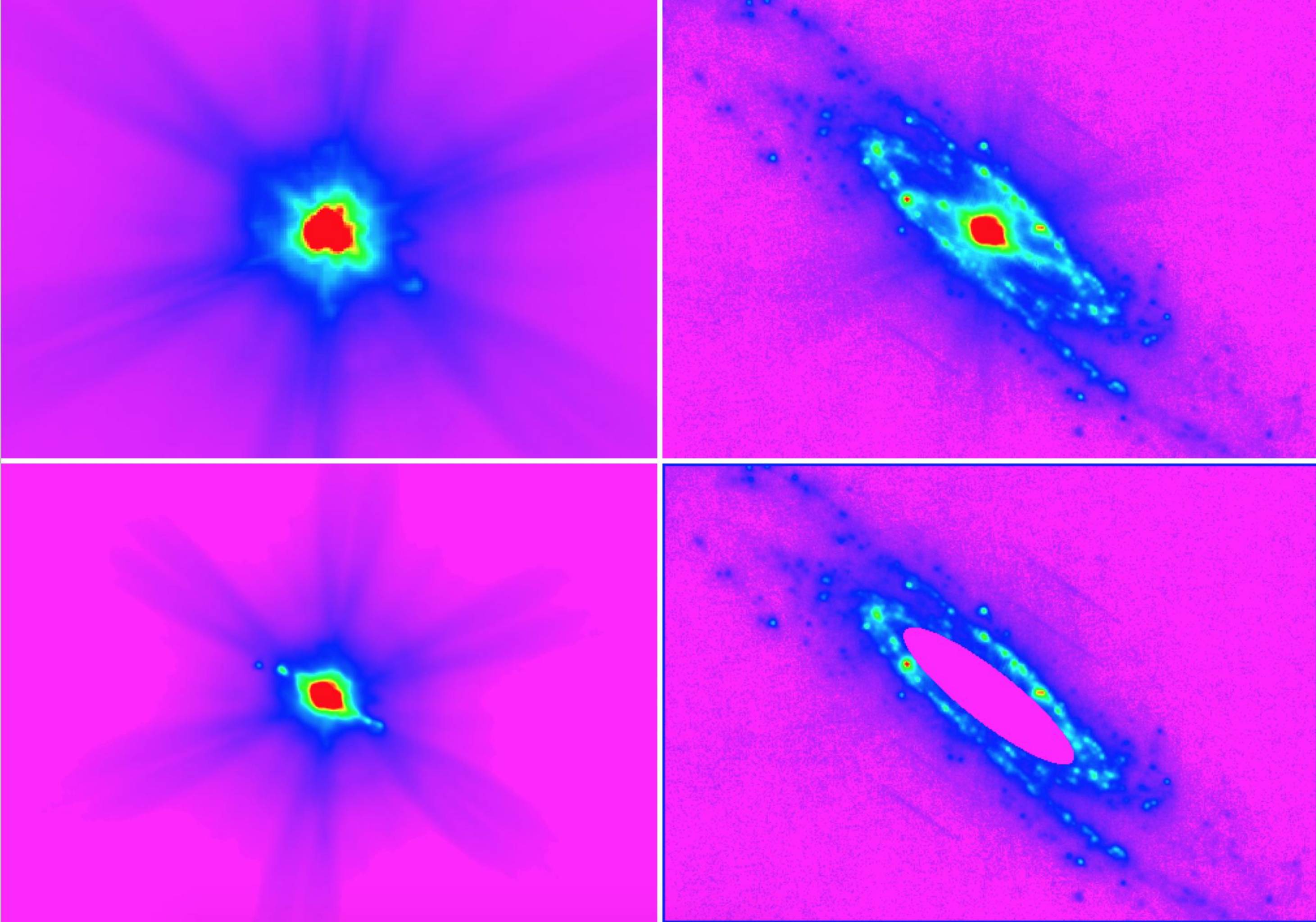}
    \caption{The procedure of PSF sidelobes in PACS $\SI{100}{\micro\metre}$ image. The top left panel is the PSF from  Bocchio et al. (2016). The top right panel shows the original image. The bottom left panel shows the re-convolution result of the inner 2.5$~\mathrm{kpc}$ region, containing both the inner disc and the lobes caused by it. The bottom right panel shows the residual image after subtraction. }
    \label{fig:lobes_removel}
\end{figure}

We use 3 of the Herschel PACS bands ($\SI{70}{\micro\metre}$, $\SI{100}{\micro\metre}$, $\SI{160}{\micro\metre}$) to study TIR distribution of NGC 253. There are two reasons for this: Firstly, Herschel has better imaging quality than other infrared (IR) observations. There are no bad pixels nor asymmetric lobes which are commonly found in Spitzer-MIPS observations. Secondly, the deep PACS PSFs measured by \cite{Bocchio2016} allow us to remove the lobes caused by the FIR emission of the extremely bright core and bar of NGC~253. Fig.~\ref{fig:lobes_removel} clearly shows four symmetric PSF lobes extending from the centre out of the disk. (There are another two lobes in parallel with the major axis, which overlap the disc and should also be removed.) To describe these lobes well, PSFs with very high dynamic range are needed. \cite{Bocchio2016} provide deep PSFs obtained in the three relevant PACS bands at different observing scanning speeds allowing a large dynamic range up to ~10$^6$; this is perfectly suitable for this project. Therefore, we divide NGC253 into two parts: the central $\sim$2.5$~\mathrm{kpc}$ region (mainly consisting of core and bars; hereafter core+bar) and the rest of the disk. Their infrared properties are summarized in Table~\ref{tab:infrared_flux}, which not only suggests that core+bar is extremely bright (contributing $\sim$60$\%$ of the TIR of the whole galaxy), but also a higher dust temperature for core+bar (the IR emission peaks at $\sim\SI{100}{\micro\metre}$) compared with the outer disk (the IR emission peaks at $\sim\SI{160}{\micro\metre}$). The extending lobes, especially those two that overlap with the disc, strongly contaminate the FIR images of the disc. As a result, they should be carefully subtracted. 

\begin{table}
	\centering
	\caption{flux density (Jy) of inner/outer regions of NGC253 at different infrared wavelength.}
	\label{tab:infrared_flux}
	\begin{tabular}{lccr} 
		\hline
		  wavelength  & core+bar & outer disc & total\\
		\hline
		$\SI{3.4}{\micro\metre}$ & 3.42 & 8.84 & 12.26\\
		$\SI{4.6}{\micro\metre}$ & 2.62 & 5.56 & 8.18\\
		$\SI{22}{\micro\metre}$ & 56.77 & 39.41 & 96.18\\
        $\SI{70}{\micro\metre}$ & 1391.00 & 402.99 & 1793.99\\
        $\SI{100}{\micro\metre}$ & 1630.73 & 989.20 & 2619.93\\
        $\SI{160}{\micro\metre}$ & 1178.56 & 1136.98 & 2315.54\\
        $\SI{250}{\micro\metre}$ & 389.85 & 635.29 & 1025.14\\
        $\SI{350}{\micro\metre}$ & 150.67 & 279.54 & 430.21\\
        $\SI{500}{\micro\metre}$ & 47.42 & 102.65 & 150.07\\
		\hline
	\end{tabular}
\end{table}

\begin{figure}
	\includegraphics[width=\columnwidth]{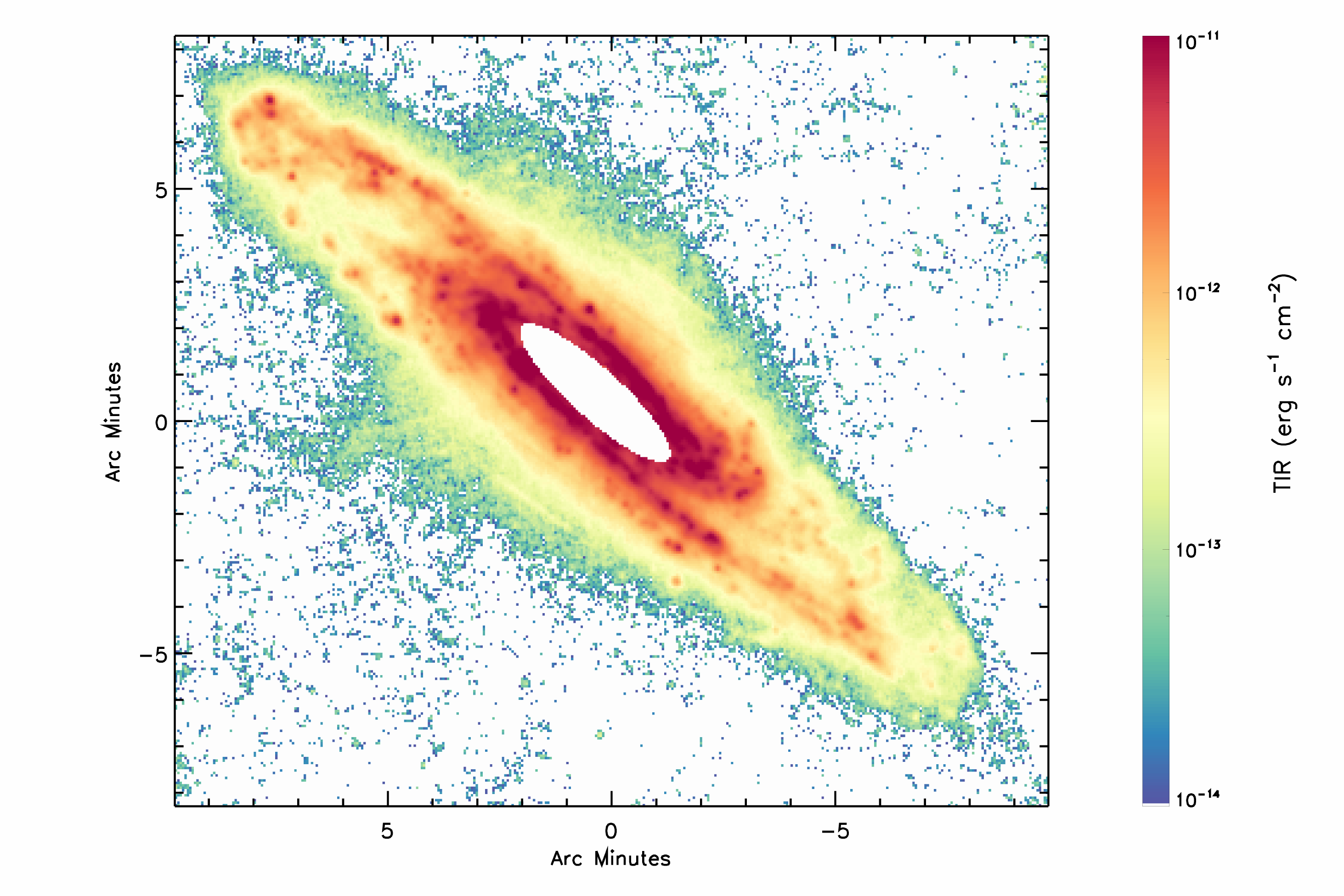}
    \caption{The TIR map of NGC253.}
    \label{fig:TIR_map}
\end{figure}

To remove these lobes, we isolate the central 2.5$~\mathrm{kpc}$ region and deconvolve it to get the position of the fluxes there. The deconvolution is implemented using the maximum likelihood deconvolution procedure MAX$\_$LIKELIHOOD\footnote{\url{https://idlastro.gsfc.nasa.gov/ftp/pro/image/max_likelihood.pro}} provided by the NASA–Goddard Space Flight Center IDL Astronomy User’s Library\footnote{\url{https://idlastro.gsfc.nasa.gov/}}. The deconvolution is performed iteratively until the image looks clean. Then the result is restored using \cite{Bocchio2016}'s PSF, trying to reproduce the six lobes as shown in figure~4 in \cite{Bocchio2016}. The PSFs we adopt are those derived from parallel scanning mode and with a scanning speed equal to 20\arcsec\ $s^{-1}$. Subsequently, both the central region and its six extending lobes are subtracted from the images. The removal procedure for the PACS $\SI{100}{\micro\metre}$ image is illustrated in Fig.~\ref{fig:lobes_removel} as an example. Similar procedures were also applied to the PACS $\SI{70}{\micro\metre}$ and $\SI{160}{\micro\metre}$ image. The top right panel is the original PACS $\SI{100}{\micro\metre}$ image, and the bottom right panel is the one after removal, which shows that most of the lobes have been removed by our method. After this procedure is carried out for all 3 Herschel/PACS bands, the $\SI{70}{\micro\metre}$ and $\SI{100}{\micro\metre}$ images were convolved to the PACS $\SI{160}{\micro\metre}$ resolution. The TIR map was calculated using the coefficients provided in Table~3 of \cite{Galametz2013}, as shown in Fig.~\ref{fig:TIR_map}.

It is also noteworthy that a data combination including Herschel/SPIRE $\SI{250}{\micro\metre}$ was recommended by \cite{Galametz2013} in calculating the TIR map for a better constraint of the sub-mm slope. But we still decided to exclude the SPIRE $\SI{250}{\micro\metre}$ data from the final TIR calculation, because it contains instrumental artifacts, especially lobes similar to those observed in the PACS data; these significantly reduce the data quality of the SPIRE $\SI{250}{\micro\metre}$ image. We could not fully remove the lobes due to the lack of a high dynamic range PSF. In addition, a TIR ($TIR_{250}$) version of the NGC~253 comprised of the SPIRE $\SI{250}{\micro\metre}$ data and the PACS $\SI{70}{\micro\metre}$/$\SI{100}{\micro\metre}$/$\SI{160}{\micro\metre}$ data was also generated and compared with the TIR image made using just the data from the three PACS bands ($TIR_{160}$); and convolved to the same resolution as the SPIRE $\SI{250}{\micro\metre}$) image. 

In most of the disc regions (except for the outer regions strongly affected by the artifacts), the differences between $TIR_{160}$ and $TIR_{250}$ are small (within 10 percent). The resulting uncertainties are similar to the SPIRE $\SI{250}{\micro\metre}$ version of NGC~628 and NGC~6946's TIR map, as shown in Figure 2 of \cite{Galametz2013}. Therefore, we adopt the TIR version using a combination of PACS $\SI{70}{\micro\metre}$/$\SI{100}{\micro\metre}$/$\SI{160}{\micro\metre}$.

Finally, FUV and H$\alpha$ images were also convolved into 11.18$\arcsec\ $ resolution (equal to the TIR map). They were corrected using the TIR map. The correction coefficients for FUV and H$\alpha$ flux are available in the Table~3 of \cite{Hao2011} and the Table~4 of \cite{Kennicutt2009}. To derive the star formation rate from the unreddened FUV/H$\alpha$ flux, we adopted the model prediction from \cite{Hao2011}, which uses the Starburst99 \citep{Leitherer1999, Vazquez2005} stellar population model under the assumption of a Salpeter IMF \citep{Salpeter1955} and a constant SFH during the past 100 Myr. The coefficients to calculate star formation rate from FUV/H$\alpha$ luminosity are summarized in Table 2 of \cite{Hao2011}. The total star formation rates traced by FUV and H$\alpha$ in NGC253's disk are 2.68 \(M_\odot\)\si[per-mode=symbol]{yr^{-1}} and 2.19 \(M_\odot\)\si[per-mode=symbol]{yr^{-1}} respectively. Thus, the two indicators agree with each other to about 20\%\ accuracy. This suggests the model assumptions are acceptable to first order. The resolved IRX-$\beta$ relation, which is the relation between the luminosity ratio of TIR and FUV (IRX), and the UV spectral slope ($\beta$) \citep{Calzetti1994, Meurer1999}, was also employed to test the dust attenuation correction, as detailed in Section~\ref{subsubsection:irx_beta}.

\section{\HI\ data analysis}
\label{section:HI}

\subsection{Rotation curve fitting}
\label{subsection:RC_fitting}
\begin{figure}
	\includegraphics[width=\columnwidth]{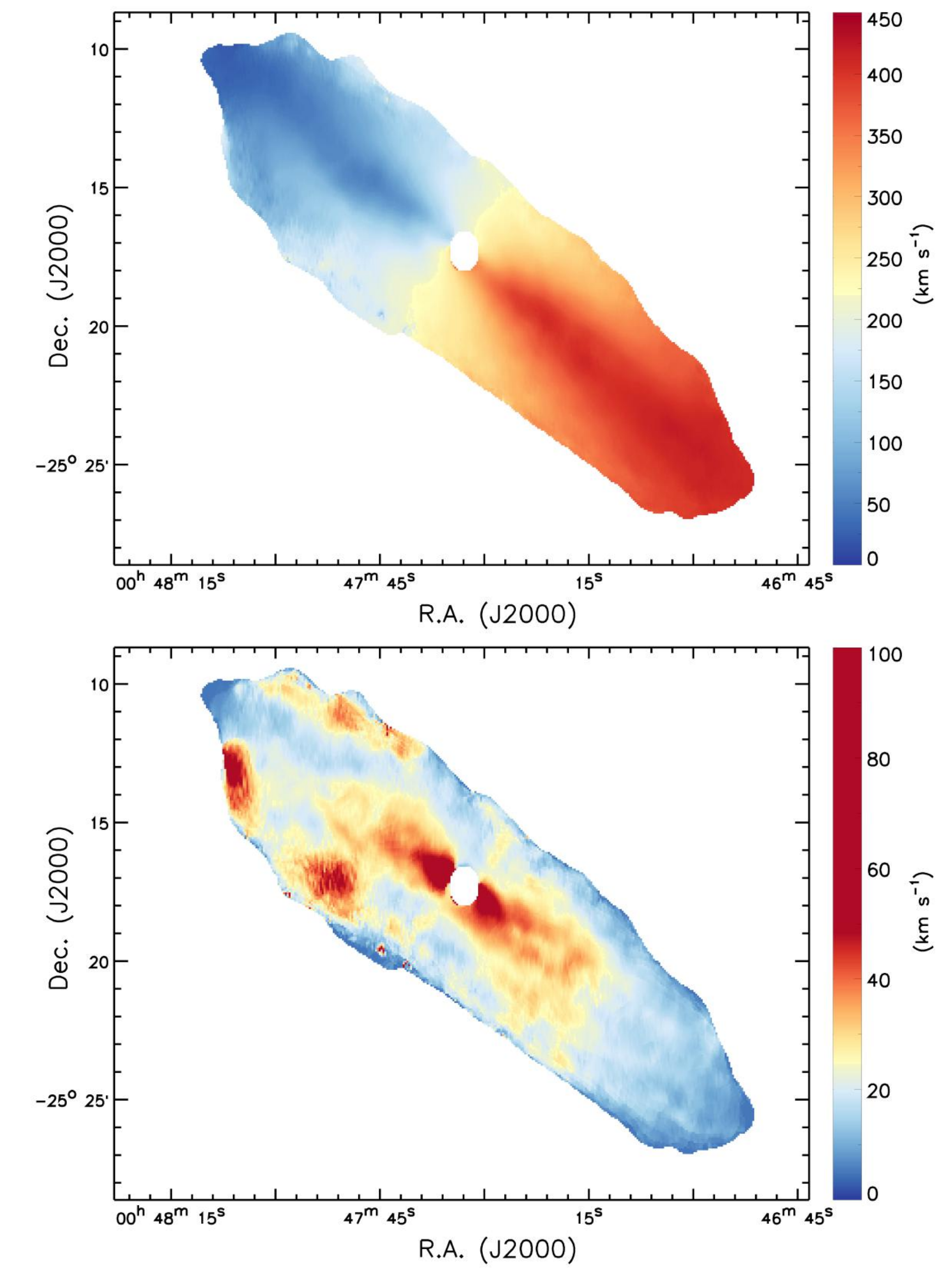}
    \caption{Velocity field (top) and velocity dispersion (bottom)}
    \label{fig:moments}
\end{figure}

\begin{figure}
	\includegraphics[width=\columnwidth]{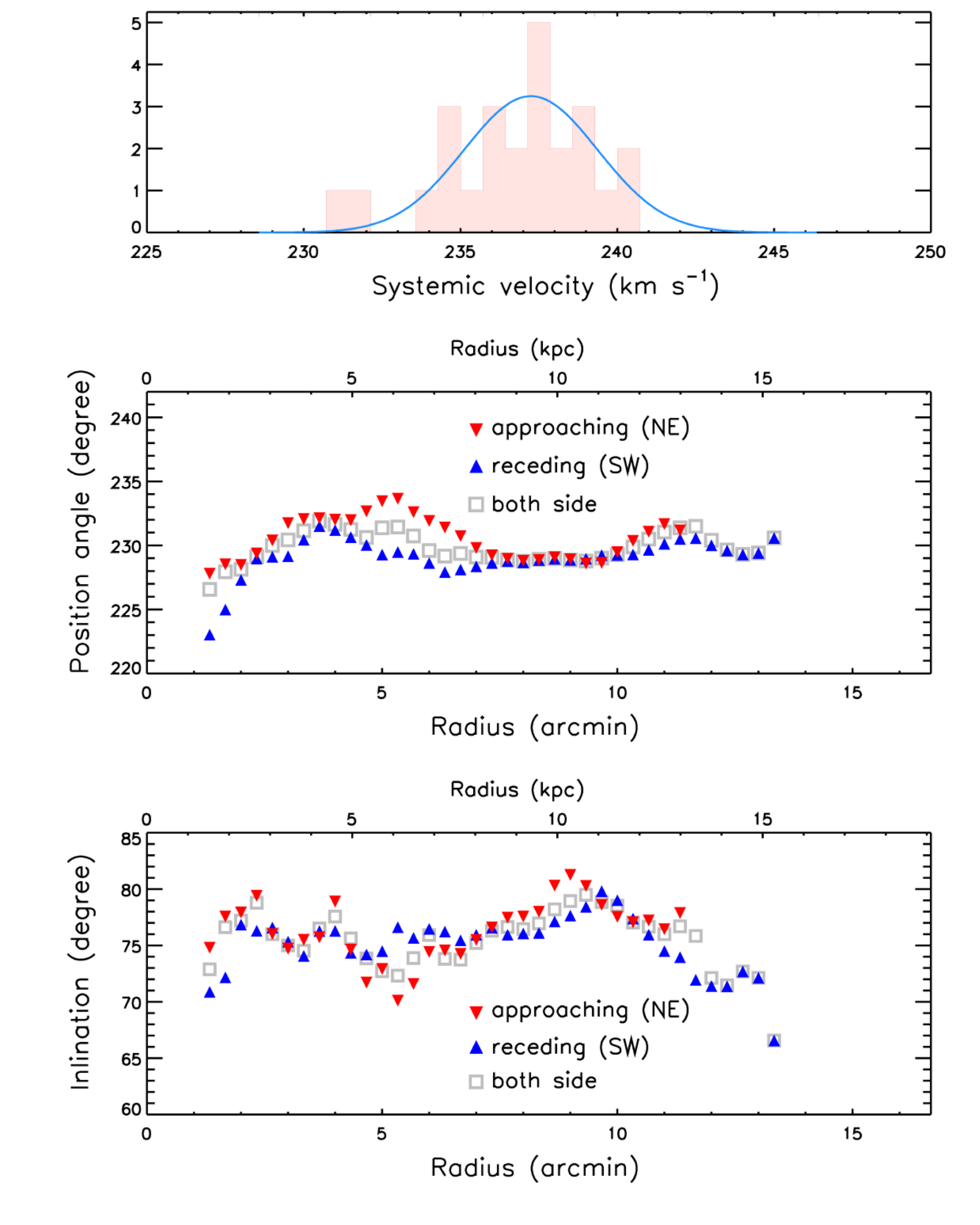}
    \caption{Top: systemic velocity, middle: position angle, bottom: inclination}
    \label{fig:vsys_pa_incl}
\end{figure}

\begin{figure}
	\includegraphics[width=\columnwidth]{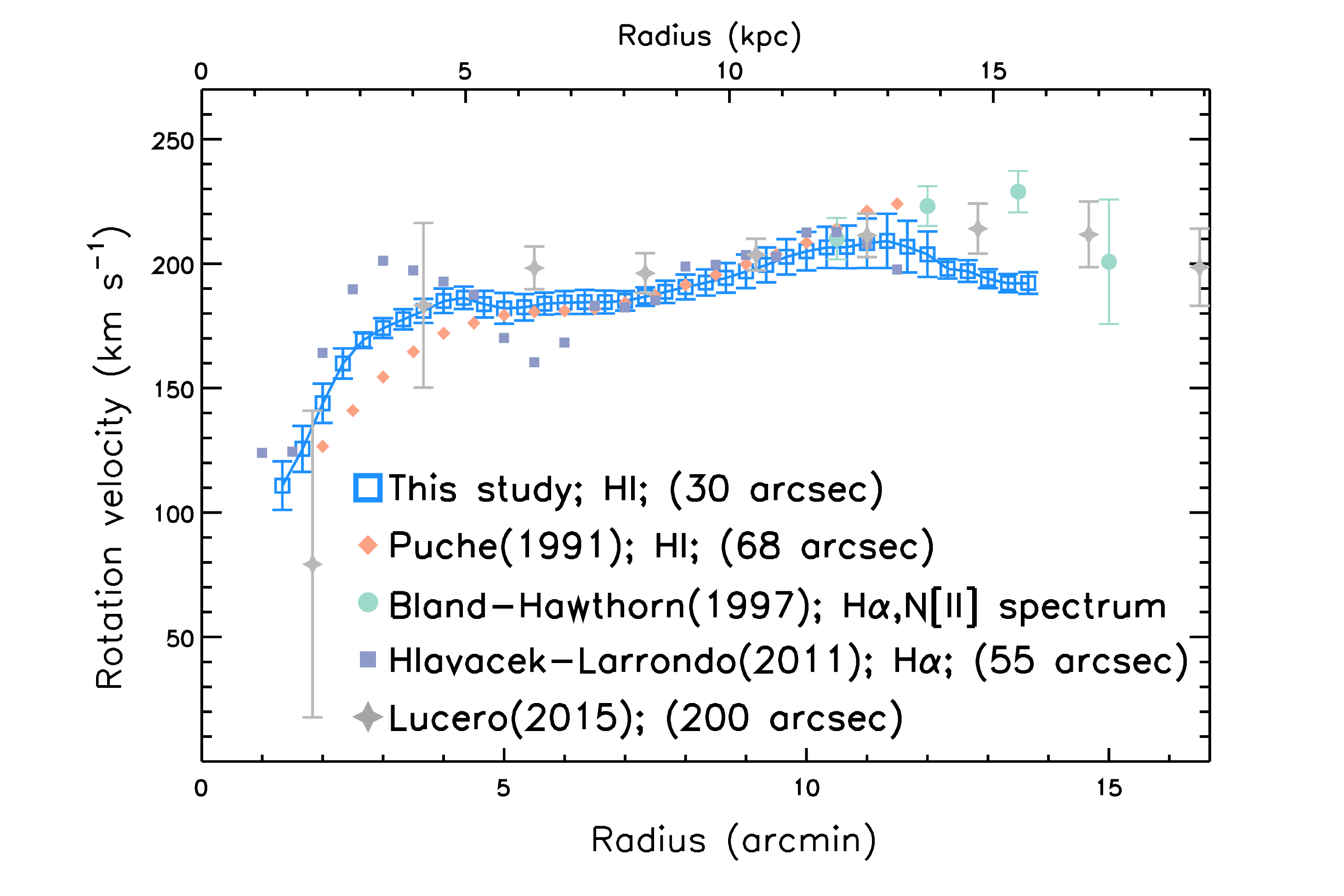}
    \caption{Derived rotation curve of NGC 253 (open, blue squares) compared to previous rotation curves from the literature.}
    \label{fig:RCs_pure}
\end{figure}

To derive the rotation curve, the high-resolution ($\sim$ 0.5 \arcmin\ = 573 pc) velocity field (top panel of Fig.~\ref{fig:moments}) was fitted with a tilted-ring model using the GIPSY \citep{van1992} task ROTCUR \citep[]{ Begeman1989}. A tilted-ring width of 20 arcsec was chosen, which is around two-thirds of the beam size. A cosine-squared weighting function and exclusion of all data points within a cone centred on the minor axis and having an opening angle of 65$^{\circ}$ (as a comparison: \cite{Lucero2015} 60$^{\circ}$; \cite{Hlavacek-Larrondo2011} 65$^{\circ}$) was adopted. 

In our tilted-ring RC analysis, we fit all the relevant parameters, which include the dynamical centre and systemic velocity, the position angle and inclination. Following \cite{Hlavacek-Larrondo2011} and \cite{Lucero2015}, two sets of parameters and the RC were obtained in three procedures. 

\begin{enumerate}
\item By fixing the position angle (PA) and inclination (INCL) at initial values from Koribalski B. S. (2018), the galaxy centre and systemic velocity were fitted for interior regions ($1\arcmin < R \leqslant 10\arcmin$) of the disc. A Gaussian function was fitted to the histogram of the results as shown in the top panel of Fig.~\ref{fig:vsys_pa_incl} to find the best-fit parameters. The rotation centre is also determined in the same way. A systemic velocity of 237.14$\pm{2.98}$ \kms\ and a rotation centre of ($00^{\si{h}} 47^{\si{m}} 33.1^{\si{s}}$, $\ang{-25;17;20.9}$) was thus found and adopted, consistent with earlier studies. 

\item The position angle and inclination were then derived (bottom two panels of Fig.~\ref{fig:vsys_pa_incl}) while the galaxy centre and systemic velocity were fixed at the values listed above. There is a PA and INCL variation similar to \cite{Puche1991} and \cite{Hlavacek-Larrondo2011}. However, the variation is fairly small except for a few data points at the very outskirts of the disc (which was probably caused by a lack of data points). As a result, the position angle and inclination were treated as a constant (PA = 239.9$^{\circ}$, INCL = 76$^{\circ}$), which is also determined by fitting a Gaussian function to their distributions and adopting its peak value.

\item Finally, the rotation curve was determined while holding all other parameters fixed to the values noted in the previous steps. This results in the RC shown in Fig.~\ref{fig:RCs_pure}.
\end{enumerate}

\subsection{Gaussian decomposition of \HI\ pixel-spectral line}
\label{subsection:Gaussian_decomp}

Both \cite{Boomsma2005} and \cite{Lucero2015} adopt the same technique to separate anomalous gas from the disc. They extract position–velocity (PV) slices that are aligned with the major axis, and visually inspect them in order to mask any emission that seems to be kinematically anomalous. However, two drawbacks of this technique will affect the accuracy of the final separation of the anomalous \HI\ from the disc. First, visual inspection and artificial masking will introduce extra uncertainties. More importantly, the accuracy of kinematical separation using PV slices decreases with increasing beam size. Meanwhile, to properly detect anomalous gas, a fairly large beam size is needed to reach a high column density sensitivity. Therefore, we attempt to solve these problems by isolating the anomalous gas in an unbiased and uniform way: for each pixel of NGC~253's data cube, we apply a Gaussian decomposition analysis to its line profile and subsequently select the non-rotational components by comparing with the modelled velocity field from RC fitting. 

To distinguish anomalous gas from rotation, multiple Gaussian components should be fitted to the \HI\ emissions of every pixel in \HI\ data cubes. We developed an IDL toolkit called FMG (Fit Multiple Gaussian components) based on the $\chi^{2}$ minimization procedure to fit multiple Gaussian components to \HI\ data cubes. Proper initial guesses for the Gaussian parameters are automatically found to avoid getting stuck in a local minimum, which is a common issue of the $\chi^{2}$ minimization technique. The number of Gaussian components adopted was determined by using the Bayesian information criterion \citep[BIC; ][]{Schwarz1978}. Details of the FMG toolkit are available in section~\ref{section:details_toolkit}. Three data sets of mock spectra were also fitted to test the reliability of our toolkit, which will be discussed in Section~\ref{subsubsection:mock_spectrum_test}. Finally, the two versions of \HI\ data cubes were fitted using our toolkit. The fitting results are discussed in Section~\ref{subsubsection:HI_fitting_result}.

\subsubsection{Test using mock \HI\ spectrum}
\label{subsubsection:mock_spectrum_test}
To test the robustness and reliability of our toolkit, 1,500 mock spectra with different velocity resolutions and different numbers of input Gaussian components were generated and fitted. The spectra are divided into three sets according to velocity resolutions. Each set contains 500 spectra with 2-6 Gaussian components. Gaussian noise was added to each spectrum. Since one of the key purposes of this project is to recognize anomalous \HI\ emission, a broad component with velocity dispersion wider than 50 \kms\ was also added to all mock spectra. The parameters of the input Gaussian components were randomly selected in a certain range, the details of which are available in Table~\ref{tab:mock_spectra_input_parameters}. The BIC value was calculated using data points above the 3 $\sigma_{rms}$ threshold. The calculation of BIC values is illustrated in Appendix~\ref{subsubsection:BIC_selection}.

\begin{figure}
	\includegraphics[width=\columnwidth]{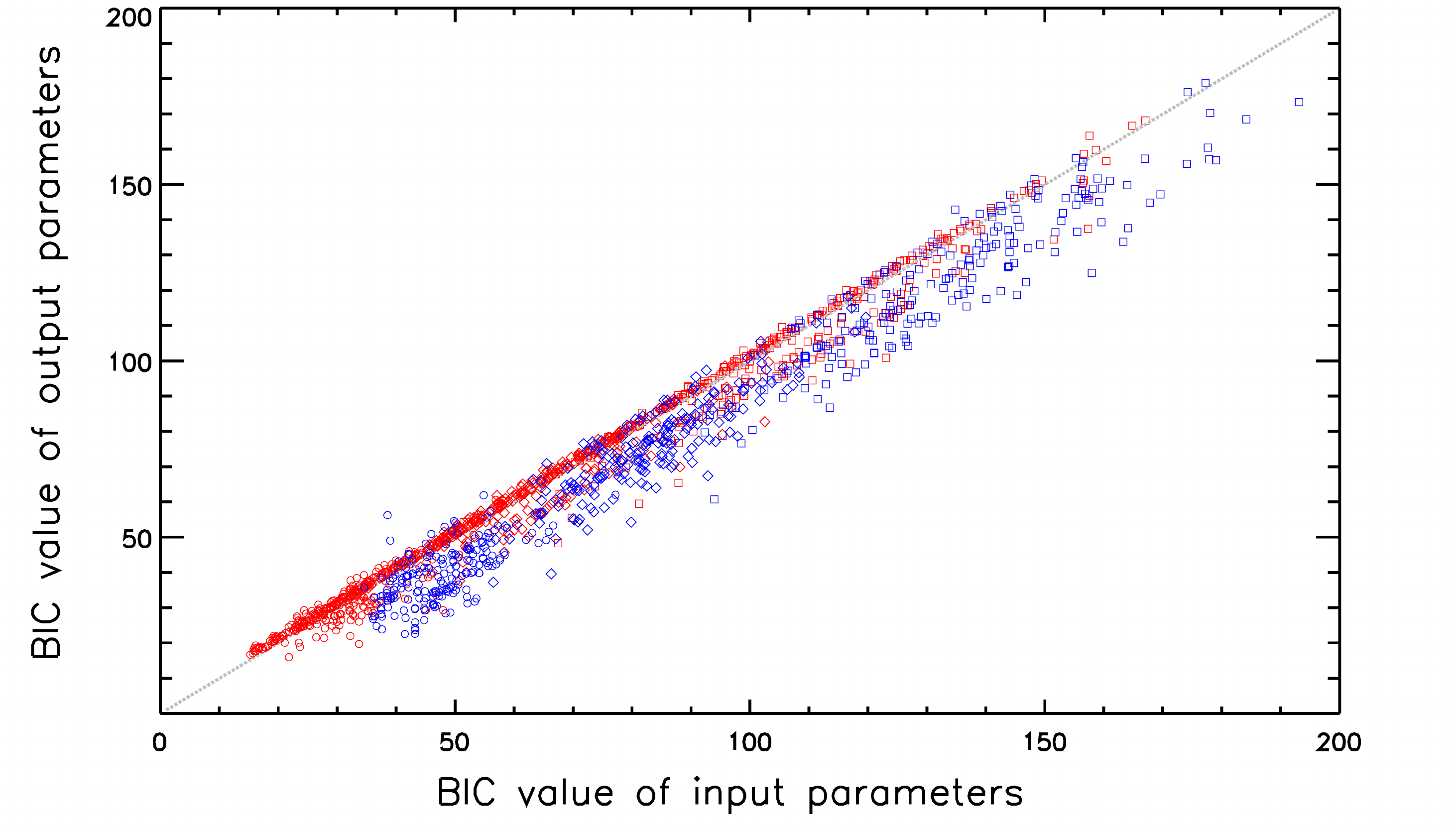}
    \caption{The BIC values from our fit versus those calculated using input parameters. Spectra with different velocity resolutions were plotted in different shapes (4 \kms: open circle, 8 \kms: open diamond, 20 \kms: open square). Spectra with 2-4 Gaussian components are plotted in red and those with 5-6 components are plotted in blue.}
    \label{fig:bic_distribution_mock}
\end{figure}

To evaluate the fitting quality, we compared the final BIC values with those calculated using input parameters, as shown in Fig.~\ref{fig:bic_distribution_mock}. It shows that almost all solutions have similar or smaller BIC values than input parameters, which not only suggests that proper solutions for all spectra were found without being trapped in local minima but also indicates that no extra Gaussian components were used compared with the input. Notably, those spectra with 5-6 components have systematically smaller BIC values than their inputs. This is caused by the degeneracy of the complex line profiles. In those cases, several Gaussian components are merged into a single component (as shown in Fig.~\ref{fig:mock_spec_G5_6}) so that a smaller number of components were needed to describe the emission line than input. More discussion on this is available in Appendix~\ref{section:details_toolkit}.

\begin{table}
	\centering
	\caption{Details of input parameters of mock spectra.}
	\label{tab:mock_spectra_input_parameters}
	\begin{tabular}{lcc  } 
		\hline
		     & normal component & broad component \\
		\hline
		components per spectrum& 1-5 & 1 \\
            
		central velocity range & 220-380 & 250-350 \\

            (\kms\ )  &  &   \\

		velocity dispersion range & 10-30 & 50-100 \\

            (\kms\ )  &  &   \\
		peak value range & 20-100 & 5-30 \\

            ($\sigma_{rms}$)  &  &   \\
		\hline
	\end{tabular}
\end{table}

\begin{figure}
\includegraphics[width=\columnwidth]{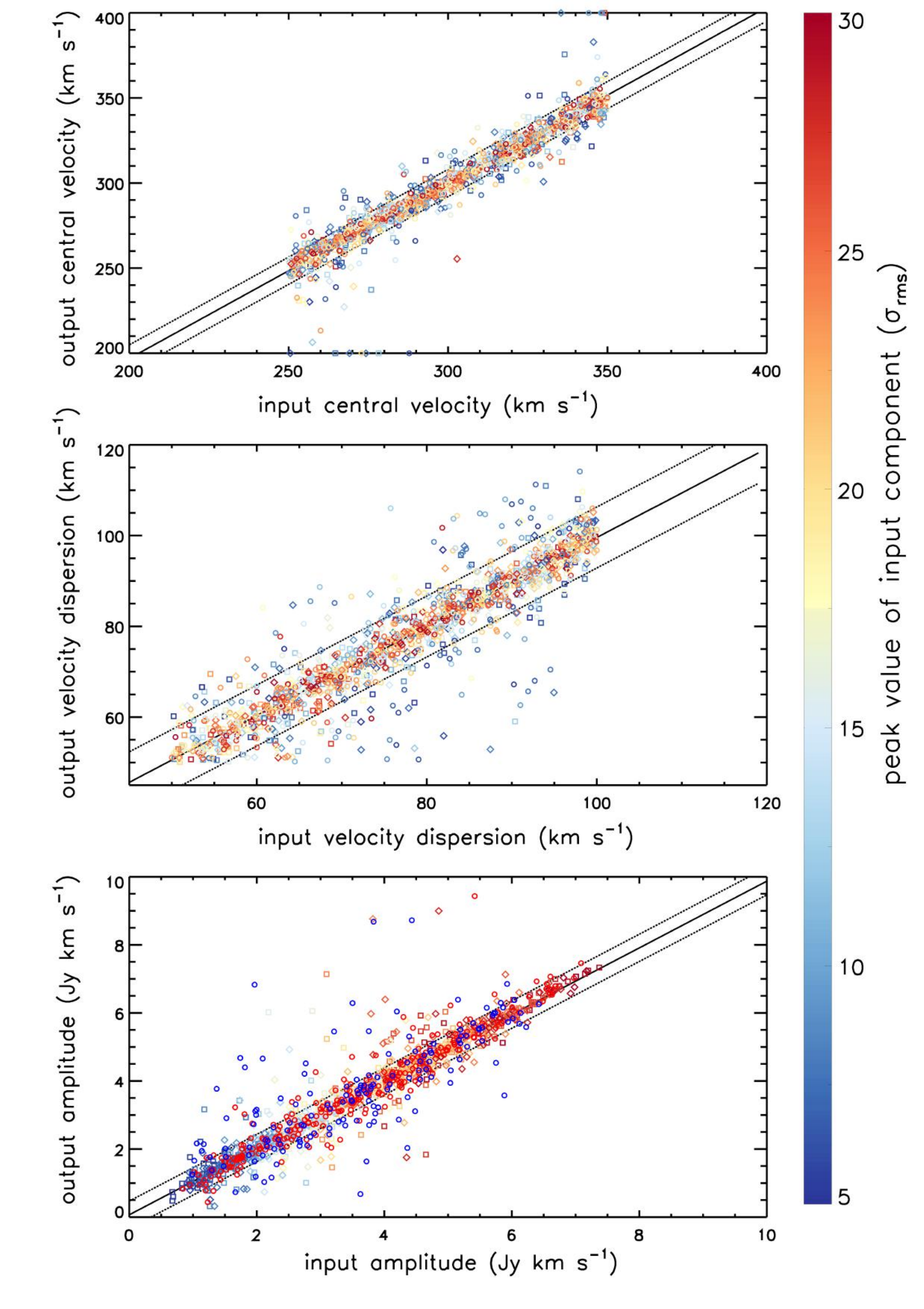}
    \caption{Input parameters versus output parameters of broad components in mock spectra. Different shapes have the same meanings as in Fig.\ref{fig:bic_distribution_mock} (4 \kms\  : open circle, 8 \kms\  : open diamond, 20 \kms\  open square). Components are colourized according to their peak value as shown by the colour bar on the right side. The fitting result from linear regression is plotted as the black lines and the 1 $\sigma$ scatter is shown as the dashed lines.}
    \label{fig:fitting_results_anomalous_mock}
\end{figure}

Another way of estimating the robustness of the fitting results is to compare the Gaussian parameters with the input ones. Table~\ref{tab:input_output_parameters_scatter} presents the 1 sigma scatter of parameter differences between input and output parameters. The results suggest:

\begin{enumerate}
\item Our toolkit provides reliable parameter estimates for both narrow and broad kinematical components. Even in the worst situation (broad components in a complex spectrum at a resolution of 20\kms\ ) the uncertainty (16.18\kms\ ) is still smaller than the velocity resolution, which confirms the necessity to decompose different components for detailed kinematical studies.
\item Generally, parameters of spectra with lower velocity resolution are harder to estimate. Meanwhile, the complexity (number of input components) of the spectrum has a bigger effect. For example, the velocity uncertainties for narrow components in simple spectra increase from 0.55 \kms\ to 1.56 \kms\ when the velocity resolution decreases from 4 \kms\ to 20 \kms.\ But they increase to 11.88 \kms\ in complex spectra (spectrum having 5-6 components). Fortunately, this is not expected to be a problem for anomalous \HI\ in NGC~253 because most of the diffuse gas was found in the outer disc of NGC~253 where the gas is not kinematically complex.
\item The parameter uncertainties of broad components, especially the uncertainties of the central velocity and velocity dispersion, are systematically larger than those of narrow ones. However, the relative uncertainties are acceptable considering their large velocity dispersion (50-100\kms\ ), as shown in Fig.\ref{fig:fitting_results_anomalous_mock}. Additionally, components with higher peak flux density show less scatter in the parameter estimation, which is also one of the reasons for adopting a peak threshold for anomalous \HI\ subtraction in section~\ref{subsection:anmls_subtraction}.
\end{enumerate}

\begin{table}
	\centering
	\caption{Uncertainties (1 $\sigma_{rms}$) of Gaussian parameters estimation.}
	\label{tab:input_output_parameters_scatter}

    \begin{tabular}{p{0.15cm}ccccccc}
        \hline
        & & & narrow$^{a}$ & & & broad$^{a}$ & \\
        
        G.s$^{b}$& & 2-4 & 5-6 & 2-6 & {2-4} & {5-6} & 2-6 \\ 
        \hline
        V$^{c}$ & $4^{d}$ \kms\ & {0.55} & {2.81} & 0.77 & {4.27} & {9.81} & 5.31 \\ 
        
         & $8^{d}$ \kms\ & {0.82} & {3.75} & 1.29 & {4.77} & {9.21} & 6.30 \\  
         
         & $20^{d}$ \kms\ & {1.56} & {11.88} & 3.96 & {9.94} & {16.18} & 11.51 \\ 
         
        S$^{c}$ & 4 \kms & {0.91} & {3.19} & 1.61 & {3.87} & {7.40} & 5.07 \\ 
        
         & 8 \kms\ & {1.08} & {4.02} & 2.18 & {5.51} & {8.48} & 6.09 \\ 
         
        & 20 \kms\ & {2.41} & {11.00} & 4.64 & {8.34} & {14.02} & 9.61 \\ 
        
        A$^{c}$ & 4 \kms\ & {0.21} & {2.50} & 0.36 & {0.26} & {0.55} & 0.30 \\  
        
        & 8 \kms\ & {0.25} & {2.81} & 0.35 & {0.28} & {0.81} & 0.39 \\  
        
        & 20 \kms\ & {0.40} & {3.94} & 2.42 & {0.46} & {1.16} & 0.55 \\ \hline
    \end{tabular}

    Notes. $^{a}$ Narrow refers to components with velocity dispersion 10-30\kms\ , broad refers to those with velocity dispersion >50\kms\ 
      
      $^{b}$ Numbers of input Gaussian components. The scatter of parameters was calculated for three subgroups of data: simple spectra with 2-4 input components, complex ones with 5-6 input components, and all spectra.
      
      $^{c}$ V: central velocity(\kms\ ); S: velocity dispersion(\kms\ ); A: amplitude (Jy \kms\ ) 
      
      $^{d}$ Spectra with different velocity resolutions.

\end{table}

\subsubsection{Gaussian decomposition result of \HI\ data cubes}
\label{subsubsection:HI_fitting_result}

Finally, we use the toolkit to decompose the \HI\ data cubes. We set boundaries (8-200 \kms\ ) for velocity dispersion fitting since we believe that any \HI\ emission narrower than 8 \kms\  is not reasonable for NGC~253 and 200 \kms\  is wide enough as an upper limit considering the velocity range of NGC~253 (0-500 \kms\ ). All $\chi^{2}$ and BIC values were calculated using data points within the source mask produced by SoFiA. 

\begin{figure}
	\includegraphics[width=\columnwidth]{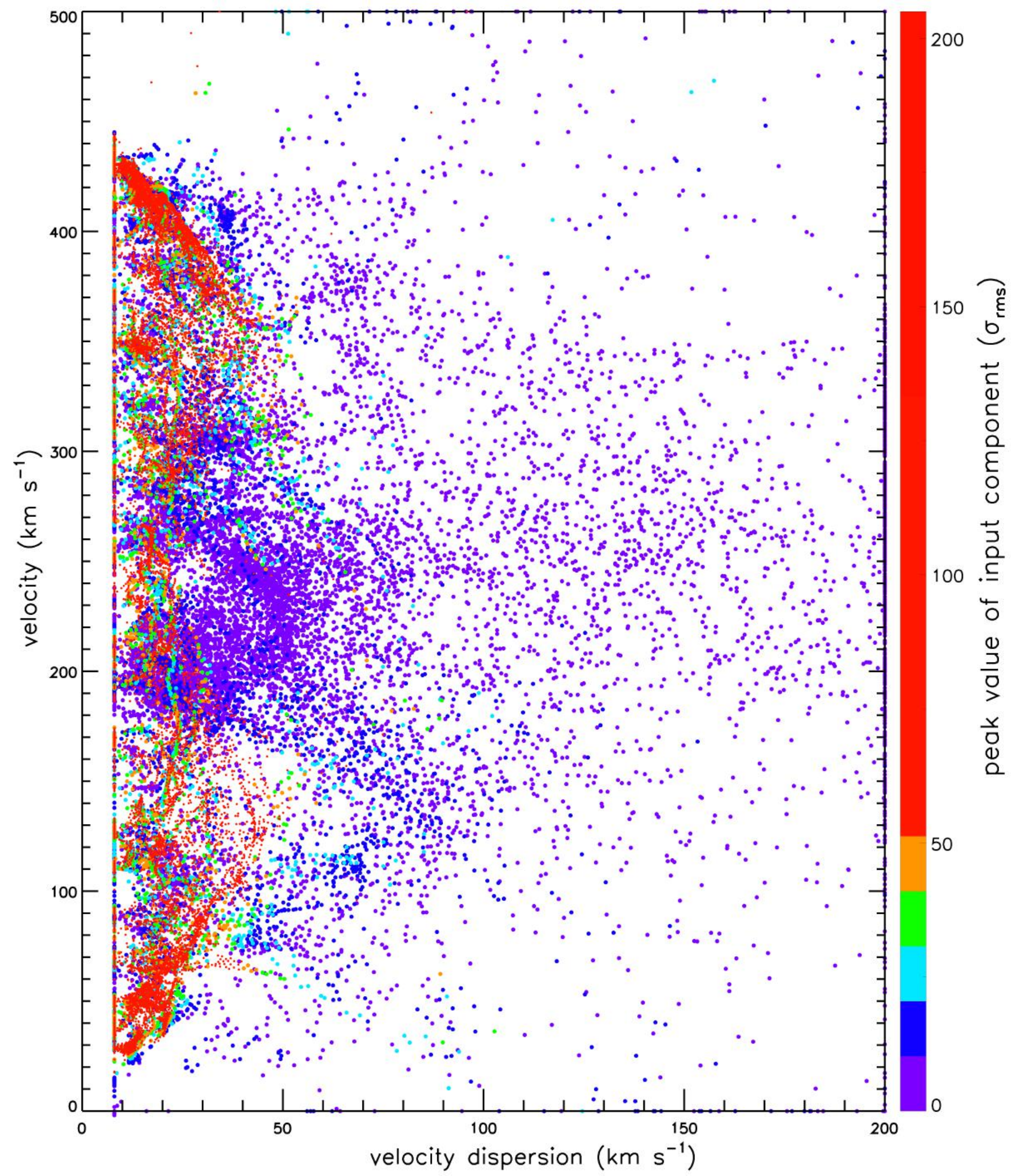}
    \caption{Velocity versus velocity dispersion plot of all kinematical components in the low-resolution data cube (resolution: 20 \kms\ ). The colours of the data points refer to their peak values as shown in the colour bar.}
    \label{fig:chi_velsigma_hi}
\end{figure}

\begin{figure}
	\includegraphics[width=\columnwidth]{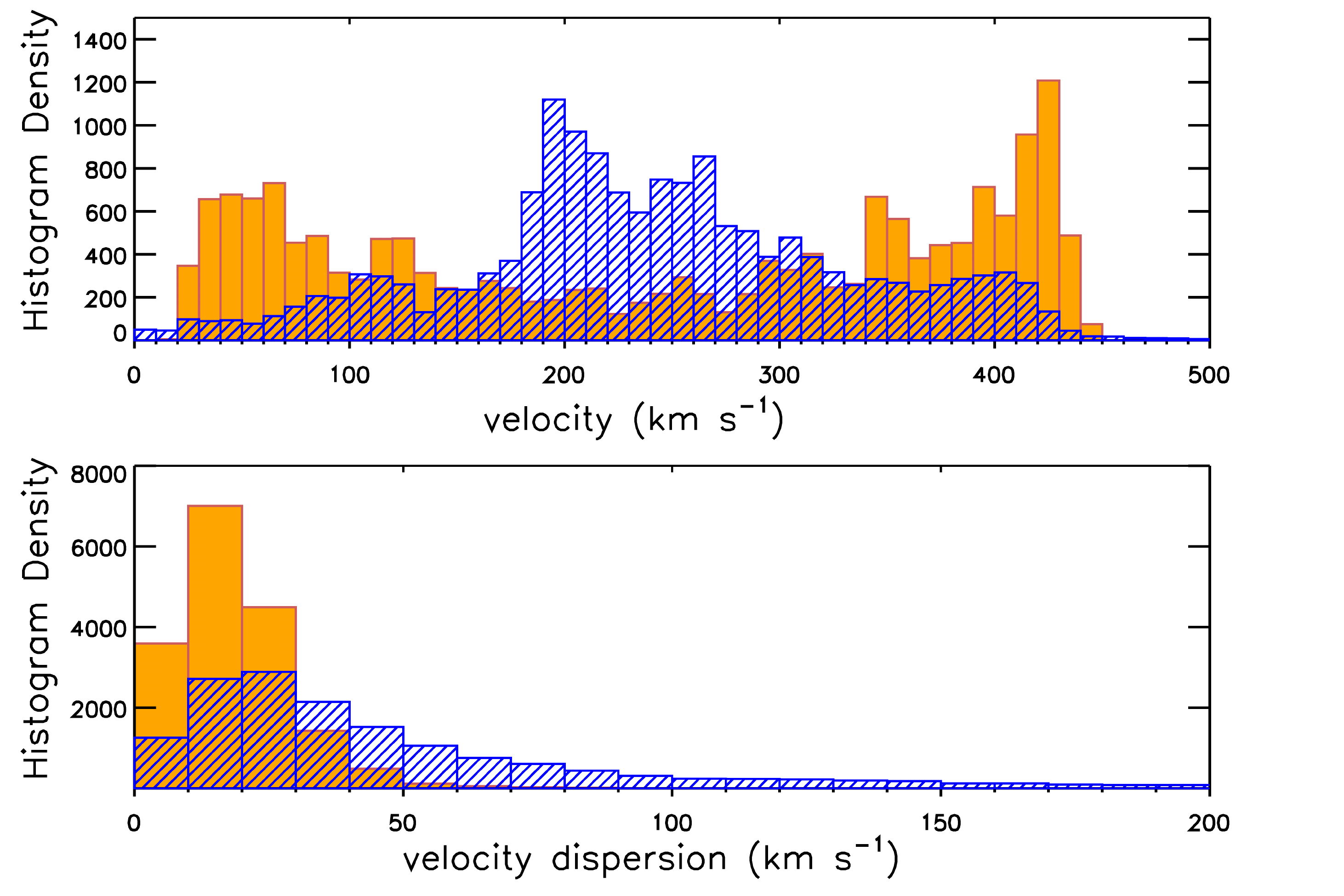}
    \caption{Top panel: Histogram of the velocity distribution of high-peak components (orange; peak value > 20$\sigma_{rms}$) and low-peak ones (blue lines; peak value < 20$\sigma_{rms}$) of the low-resolution data cube. Bottom panel: Histogram of velocity dispersion of the two groups of components.}
    \label{fig:vel_sig_hist}
\end{figure}

Fig.~\ref{fig:chi_velsigma_hi} shows central velocity versus velocity dispersion of all Gaussian components resolved in the low-resolution data cube. The vertical feature at around 10 \kms\ is caused by the lower boundary for velocity dispersion fitting. All components can be roughly divided into two groups:

\begin{enumerate}
\item The high-peak components (peak value > 20$\sigma_{rms}$) clearly show the rotation signature of the \HI\ disc, the velocity dispersion of which is mainly smaller than 40 \kms\ . The double-horn feature at $\sim$ 350 and 120 \kms\  of high-peak components is caused by the beam smearing effect, which enlarges the velocity dispersion.
\item In addition to the disc, there are many components with low peak values (peak value < 20$\sigma_{rms}$; mainly consisting of blue and purple dots in Fig.~\ref{fig:chi_velsigma_hi}) following significantly different distributions from the high-peak components. They have larger velocity dispersion than high-peak components (bottom panel of Fig.~\ref{fig:vel_sig_hist}). Also, their central velocity is mostly located between 180-280 \kms\  (top panel of Fig.~\ref{fig:vel_sig_hist}). The similar central velocity with the anomalous \HI\ found by \cite{Lucero2015} implies that the low-peak components are essentially the anomalous \HI\ we are looking for. 

\end{enumerate}

Additionally, both Fig.~\ref{fig:chi_velsigma_hi} and Fig.~\ref{fig:vel_sig_hist} show that the parameter space of rotating components overlaps with the parameter space of gas that does not rotate regularly (hereafter non-rotational gas). This indicates that a simple cut of Gaussian parameter space would not separate the anomalous gas from the rotating disc. Additional criteria taking advantage of the rotation curve should be introduced to accomplish this subtraction, which will be discussed in the next section.

\subsection{Subtraction of non-rotational gas from the \HI\ disc}
\label{subsection:anmls_subtraction}

\begin{figure}
	\includegraphics[width=\columnwidth]{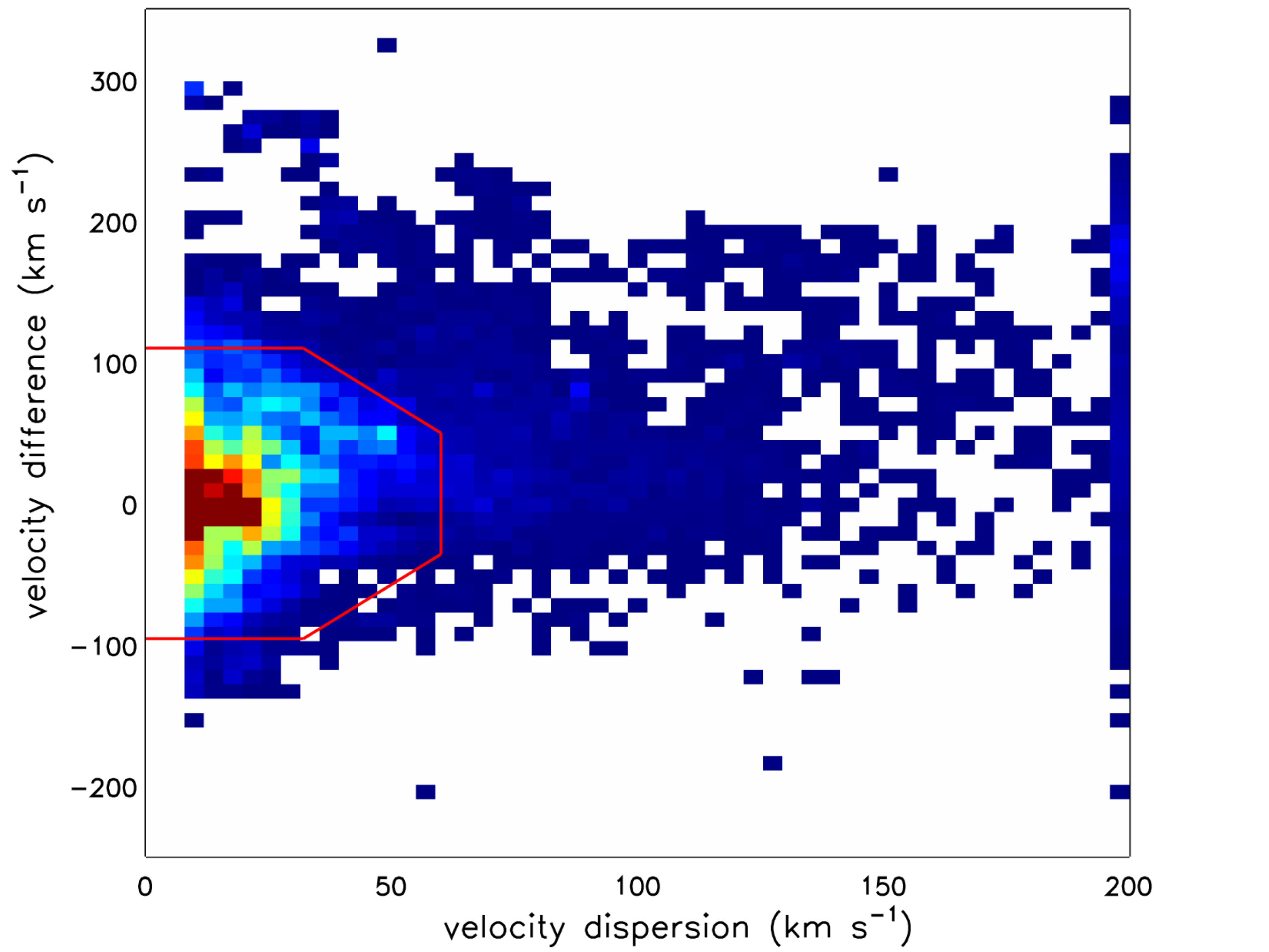}
    \caption{Density map of velocity difference versus velocity dispersion of all Gaussian components resolved in the low-resolution data cube. Velocity difference refers to the difference between the central velocity of Gaussian components and the modelled velocity from rotation curve fitting. Inside the red lines are all the rotation components defined by our subtraction criteria. }
    \label{fig:densitymap_veldif_sigma}
\end{figure}

To properly distinguish the anomalous (non-rotational) gas from the \HI\ disc, we compare resolved kinematical components with the modelled velocity field, which is generated from the fitted RC in section~\ref{subsection:RC_fitting}. For each Gaussian component, the velocity difference ($\Delta$V) is calculated by comparing its central velocity and the modelled velocity in the same position. Fig.~\ref{fig:densitymap_veldif_sigma} shows $\Delta$V versus velocity dispersion ($\sigma$) distribution for all components. Two kinematical populations are clearly visible. The overdensity feature centered at $\Delta$V $\sim$ 0 \kms  corresponds to the disc rotational motion. A large number of \HI\ components distant from it represents non-rotational motion which may be caused by gas outflow or inflow. They occupy an extensive parameter space (|$\Delta$V| up to $\sim$400 \kms, $\sigma$ up to $\sim$200 \kms). Here we propose our criteria to separate two populations:

\begin{enumerate}
\item Any component with |$\Delta$V| > 100\kms, or $\sigma$ > 60\kms, or |$\Delta$V| > 160 - $\sigma$ when 60\kms > $\sigma$ > 30\kms will be classified as non-rotational gas, as shown by the red lines in Fig.~\ref{fig:densitymap_veldif_sigma}.
\item All extra-planner gas (R greater than $\sim$19$~\mathrm{kpc}$; the change of disc shape due to elliptical beam has also been taken into account) will be recognized as non-rotational components.
\end{enumerate}

Also, the distribution of non-rotational components in Fig.~\ref{fig:densitymap_veldif_sigma} is noticeably asymmetric. There are excessive non-rotational components with $\Delta$V > 0, implying that the anomalous gas is asymmetrically distributed in NGC~253.

Finally, we isolate the non-rotational \HI\ from the disc. The spatial distribution of the non-rotational gas is shown in the bottom panel of Fig.~\ref{fig:nearby_dwarf_anmls_hafuv}. Most of the anomalous \HI\ is located on the approaching side of the galaxy. A similar distribution is also found in \cite{Boomsma2005} and \cite{Lucero2015}. 
We find a anomalous \HI\ mass of 1.02 $\pm{0.12}$ $\times$ $10^{8}$ \msun\, which corresponds to 4.1 percent of total \HI\ mass. Notably, for the components within 19$~\mathrm{kpc}$ radius, we only subtract those with peak values larger than 5 $\sigma_{rms}$. Thus, this value should be treated as the lower limit.

\section{Results}
\label{section:results}

\subsection{A actively star-forming disc}
\label{subsection:star_forming_disc}

\subsubsection{Resolved IRX-$\beta$ relation}
\label{subsubsection:irx_beta}

\begin{figure*}

    \includegraphics[width=\textwidth]{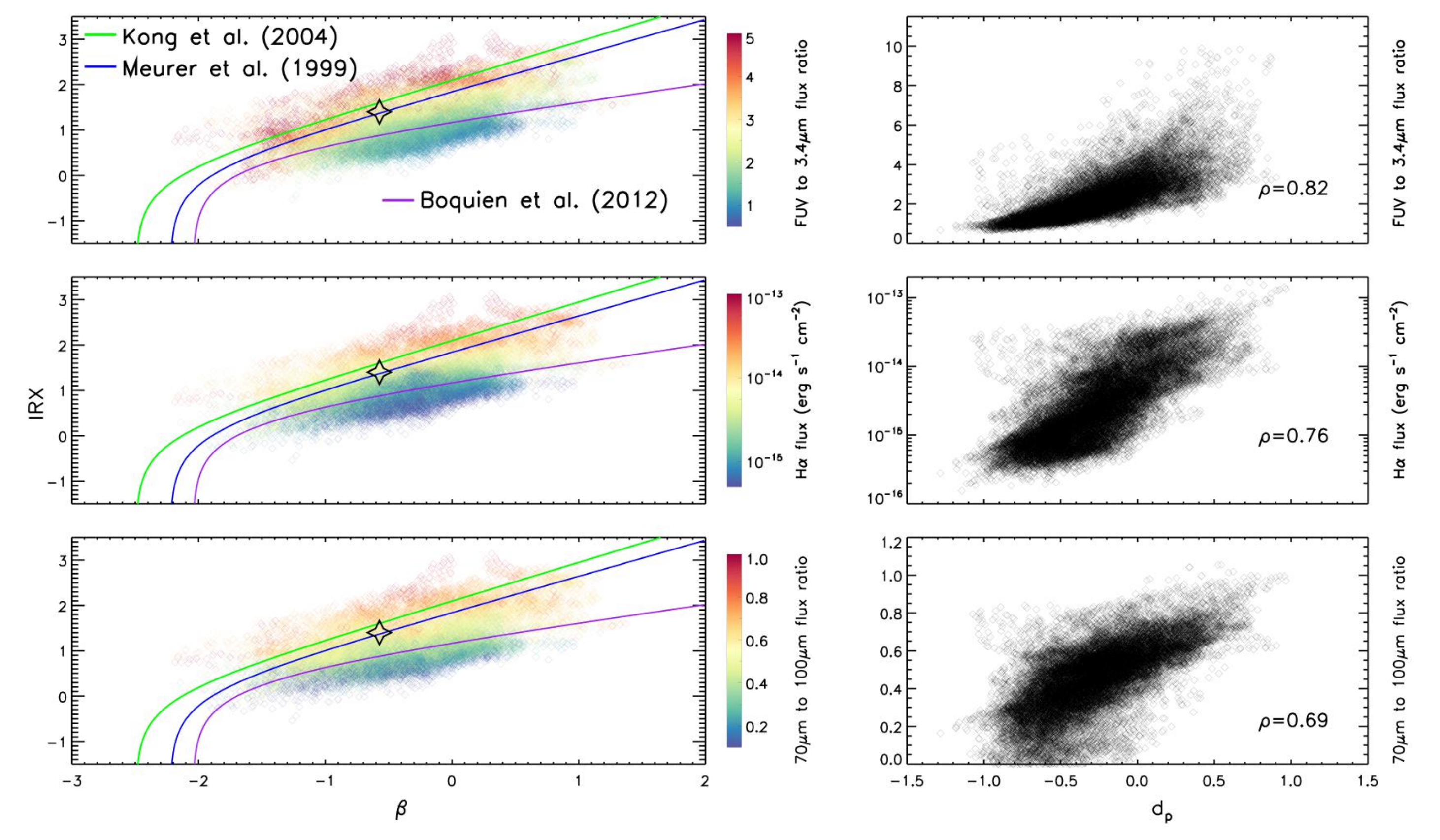}
    \caption{Left panels: IRX-$\beta$ diagram colourized in different properties with their colour bars shown at the right side. the green and blue lines represent the relation for starburst galaxies from Kong et al. (2004) and Meurer et al. (1999) respectively, while the relation for normal star-forming galaxies from Boquien et al. (2012) is shown in purple. left top: FUV to NIR flux ratio, left middle: \Ha\ flux, left bottom: $\SI{70}{\micro\metre}$ to $\SI{100}{\micro\metre}$ flux ratio. The right panels plot the perpendicular distance defined by Kong et al. (2004) to the relation from Meurer et al. (1999) versus different properties. The Spearman correlation coefficient $\rho$ is listed at the bottom right of each plot. }
    \label{fig:irx_beta_Pd}
\end{figure*}

The IRX-$\beta$ relation \citep{Calzetti1994, Meurer1999} is a well-proven tool for correcting dust attenuation, where IRX denotes the ratio between infrared and UV luminosity and $\beta$ refers to the rest-frame UV slope. Although it is an empirical relation developed to account for the global flux of galaxies, many former studies have been examining it on sub-galactic scales \citep[eg.][]{Munoz-Mateos2009, Boquien2012, Ye2016}. Here, we present the resolved IRX-$\beta$ relation of the NGC~253 disc to test the reliability of our dust correction. Possible effects causing variations in the shape of the attenuation curve, such as birthrate parameter and star formation rate, were also explored by analyzing the perpendicular distance (which is defined by \cite{Kong2004}) from the IRX-$\beta$ relation of starburst galaxies suggested by \cite{Meurer1999} (hereafter M99).

Following \cite{Boquien2012}, we perform a pixel-by-pixel IRX-$\beta$ analysis. The calculation of total infrared luminosity is explained in section.~\ref{subsubsection:dust}. The UV slope definition from \cite{Kong2004} was adopted. All images were convolved to the same resolution of 11$\arcsec$ (PACS-$\SI{160}{\micro\metre}$). The pixel size is 3.2$\arcsec$, which is around one-third of the spatial resolution (this ratio is similar to \cite{Boquien2012}). All pixels in the outer disc ($\geq$ 2.5$~\mathrm{kpc}$) with a signal-to-noise ratio greater than 5 in the unreddened FUV map were selected. The resulting IRX-$\beta$ diagram is shown in the left panels of Fig.~\ref{fig:irx_beta_Pd}. The relations from former studies are also overplotted (starburst galaxies: M99 in blue and \cite{Kong2004} in green; normal star-forming galaxies from \cite{Boquien2012} in purple). 

The global flux IRX-$\beta$ of NGC~253's outer disc is plotted as an open star. Overall, the disc can be well described by the IRX-$\beta$ relation for starburst galaxies, which suggests that the star formation activities are very intensive there. Further inspection reveals a correlation between the deviation from M99 and other intrinsic properties of NGC~253's disc, which is implied by the colour gradients in the IRX-$\beta$ diagrams. To quantify these correlations, the perpendicular distance d$_{p}$ to M99 relation was calculated for each data point. The definition of d$_{p}$ is introduced by \cite{Kong2004}, which is the shortest distance to the relation. We find that stellar age, star formation activity, and dust temperatures are tightly correlated with d$_{p}$:

\begin{enumerate}
    \item Many former studies suggest that the stellar age plays an important role in explaining the scatter of the IRX-$\beta$ relation. Following \cite{Grasha2013}, we use the unreddened FUV to NIR ($\SI{3.6}{\micro\metre}$ observation from WISE-1 band) flux ratio as the estimator of mean stellar age. The top two panels of Fig.~\ref{fig:irx_beta_Pd} show a tight correlation between FUV/NIR ratio and d$_{p}$, indicating that the regions of younger stellar age are more likely to enter the bursty mode in IRX-$\beta$ relation.
    
    \item In the middle panels, the unattenuated H$\alpha$ flux was selected as the additional parameter. The correlation is also significant. It suggests that regions with higher d$_{p}$ have more active star formation traced by H$\alpha$ emission, which is reasonable. Meanwhile, it also indicates that the dust correction of H$\alpha$ is consistent with FUV.
    
    \item The bottom two panels show the possible correlation between d$_{p}$ and $\SI{70}{\micro\metre}$/$\SI{100}{\micro\metre}$ flux ratio, which is a commonly used proxy for the dust temperature. The Spearman correlation coefficient is also large, suggesting that the bursty regions in IRX-$\beta$ relation tend to have higher dust temperatures. Meanwhile, some evidence implies that far infrared emission with $\lambda\ \leq$ $\SI{160}{\micro\metre}$ originates from dust heated by star-formating regions \citep[]{{Bendo2015_2}}. Not only does this correlation show the reliability of our dust correction results, but it also supports the idea that the dust temperature at $\lambda\ \leq$ $\SI{160}{\micro\metre}$ could be elevated by star formation activity.
\end{enumerate}

Noticeably, the Spearman correlation coefficients of our work are larger than in former studies \citep[eg.][]{Boquien2012, Ye2016}. This is probably caused by two effects. Firstly, we have removed the central core and bar in our analysis, which could have introduced additional deviations since they are in a much more active mode. Also, the physical size (~60 pc) of our data points is much smaller than in previous studies ( \cite{Boquien2012}: > 659pc; \cite{Ye2016}: \HII\ regions $\sim$ 200pc), which prevents possible contamination from the blending of different star-forming regions.

\subsubsection{Star formation activity and its correlation with other properties}
\label{subsubsection:sfr_and_other_properties}

\begin{figure}
	\includegraphics[width=\columnwidth]{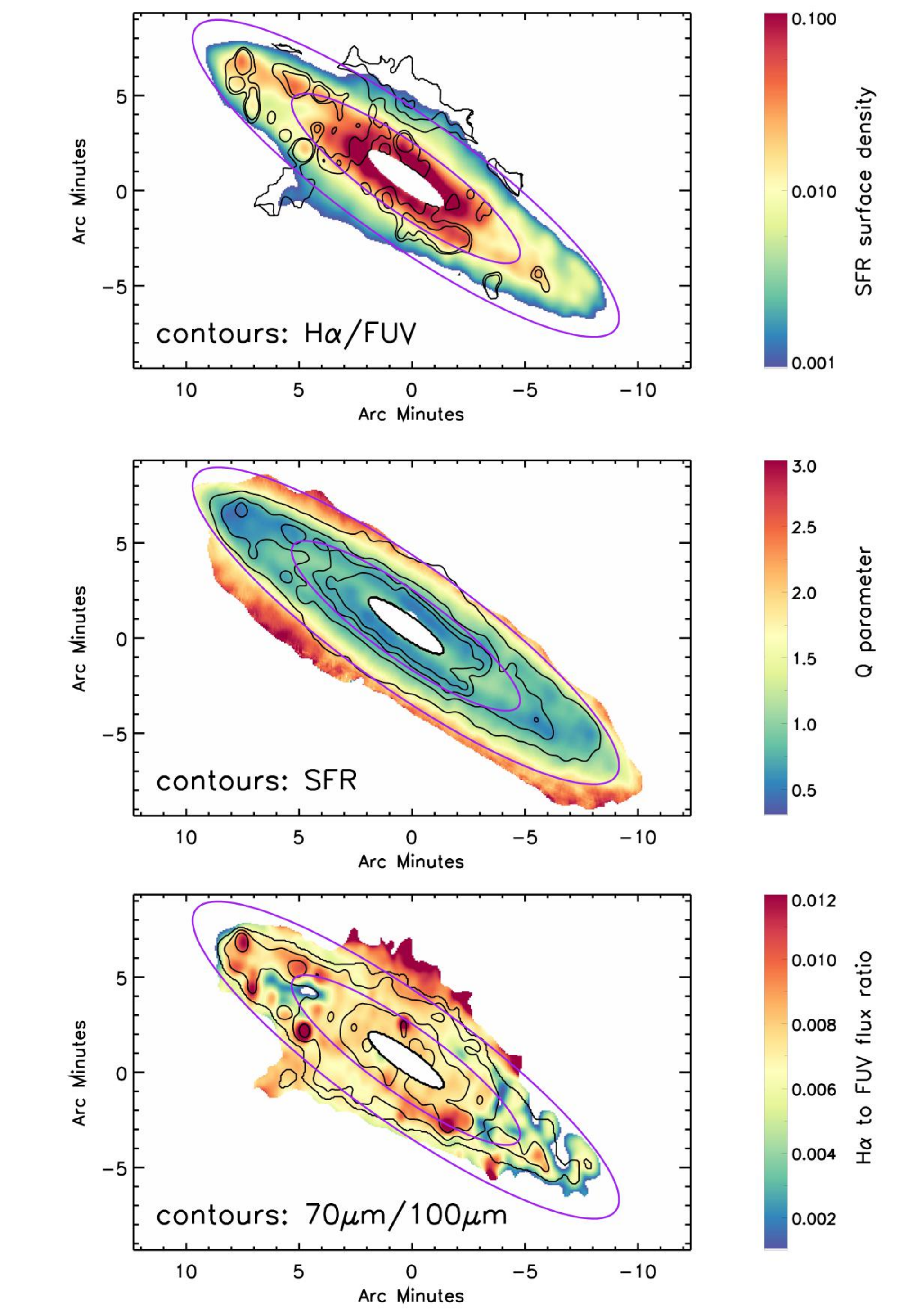}
    \caption{Top: SFR$_{FUV}$ surface density (in \msun\ yr$^{-1}$ kpc$^{-2}$) map. H$\alpha$/FUV flux ratio map is overplotted as contours (0.0065, 0.0075, and 0.0085). Middle: the disc stability parameter (with finite disc thickness) from Romeo \& Wiegert (2011) Q$_{RW,thick}$. SFR$_{FUV}$ surface density is overplotted. contours are 0.002, 0.01, 0.025 and 0.05 \msun\ yr$^{-1}$ kpc$^{-2}$. bottom: H$\alpha$/FUV flux ratio map. $\SI{70}{\micro\metre}$/$\SI{100}{\micro\metre}$ flux ratio map is presented as contours (0.1, 0.4, 0.6, and 0.8). Two purple line shows the radius of 6.5\arcmin\ and 14\arcmin,\ respectively, which are the boundaries of regions having asymmetric RC.}
    \label{fig:sfr_q_fluxratio}
\end{figure}

\begin{figure*}
	\includegraphics[width=\textwidth]{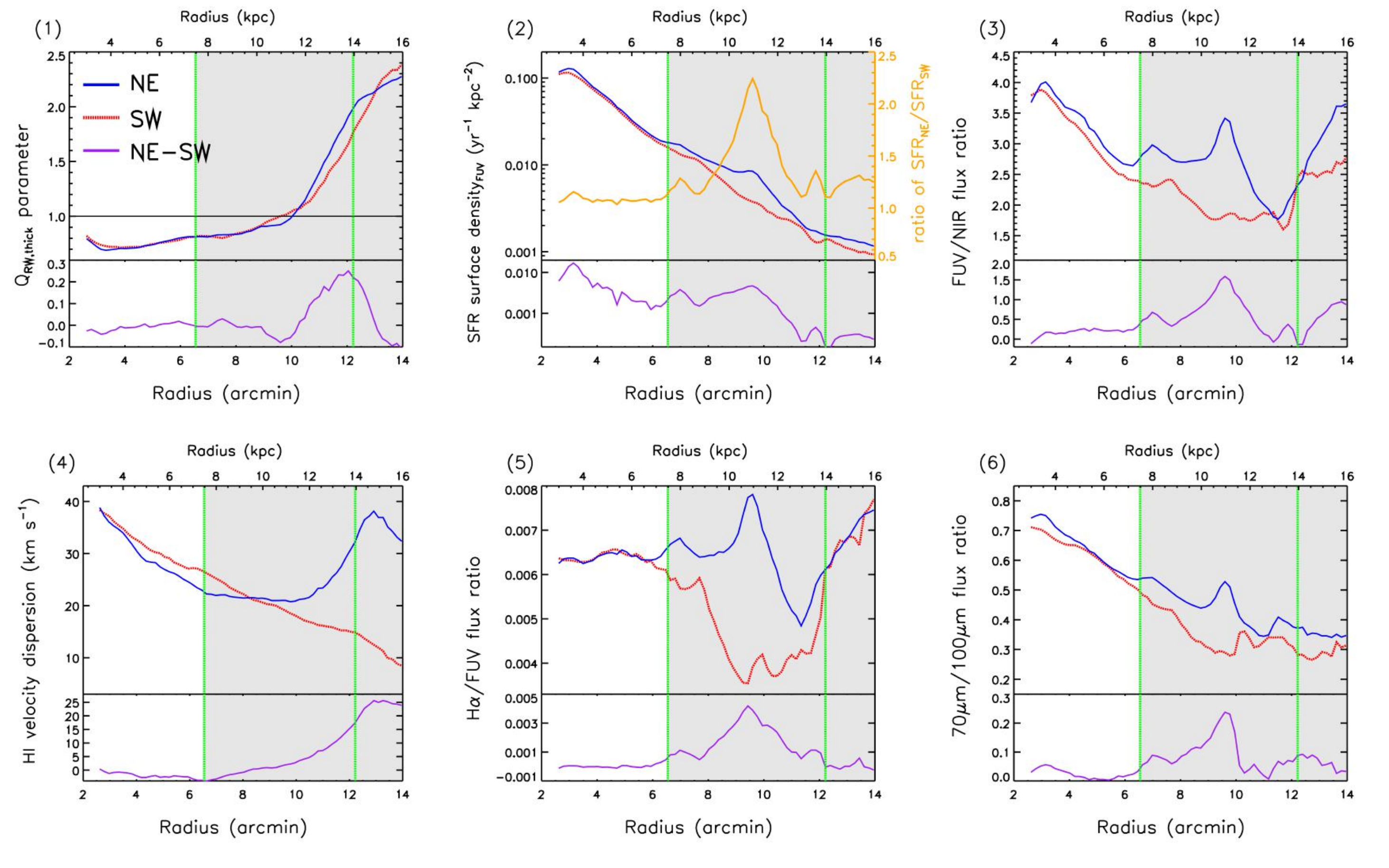}
    \caption{radial profile (3--16$~\mathrm{kpc}$) of different properties in the northeast half (approaching side; blue line) and the southwest half (receding side; red dashed line). the difference between the NE and SW is also plotted at the bottom of each panel. The grey background has the same meaning as Fig.~\ref{fig:RCs_both_sides}, which is the radius range (7.5--16$~\mathrm{kpc}$) of asymmetric RC and is explained in section~\ref{subsection:declining_rc}. As discussed in section~\ref{subsubsection:sfr_and_other_properties}, the disc is divided into 3 parts according to the asymmetries. Their boundaries are marked using green vertical lines. (1) disc stability parameter; (2) star formation rate surface density (derived from the unreddened FUV fluxes) with y-axis plotted in log scale, the ratio between SFR$_{NE}$ and SFR$_{SW}$ is overplotted in orange with its annotation showing on the right side; (3) FUV (unattenuated) to NIR (WISE $\SI{3.6}{\micro\metre}$) flux ratio; (4) intensity-weighted velocity dispersion; (5) unreddened flux ratio between H$\alpha$ and FUV; (6) $\SI{70}{\micro\metre}$/$\SI{100}{\micro\metre}$ flux ratio.}
    \label{fig:radial_profile_6panels}
\end{figure*}

After testing the reliability of dust correction using the IRX-$\beta$ diagram, the SFR (traced by FUV emission) surface density is shown in the top panel of  Fig.~\ref{fig:sfr_q_fluxratio}, suggesting a very active star-forming disc. Most parts of the disc have SFR surface densities ($\Sigma_{SFR}$) higher than 0.01 \msun\ $yr^{-1} kpc^{-2}$, and the inner part is especially active (greater than 0.1 \msun\ $yr^{-1}$ $kpc^{-2}$). The H$\alpha$/FUV flux ratio map is overplotted as contours, and the coloured version of the flux ratio map is shown in the bottom panel. Three trends of the spatial distributions of H$\alpha$ and FUV emission are clearly shown:

\begin{enumerate}
    \item Compared with the FUV emission, the H$\alpha$ is more concentrated as the shapes of several \HII\ regions (mostly seen on the northeast side) are still visible in the H$\alpha$/FUV flux ratio map (bottom panel) even at a resolution of $\sim30\arcsec$. Also, the H$\alpha$/FUV ratio is spatially correlated with dust temperature traced by the $\SI{70}{\micro\metre}/\SI{100}{\micro\metre}$ ratio (contours in the bottom panel). 
    
    \item A systematically higher H$\alpha$/FUV flux ratio on the northeast side of the disc is visible, which suggests that the H$\alpha$ emission is more asymmetric than the FUV. (Obviously, this does not mean that the FUV emission is evenly distributed between two sides of the disc.)
    
    \item Extraplanner patterns are observed in both the SFR density map and H$\alpha$/FUV flux ratio map, which is probably caused by the star formation triggered outflow. 
\end{enumerate}

To explore the reasons for strongly elevated star formation in NGC~253's disc, we also calculate the two fluid gravitational stability parameter, which takes the influences from both gas and stars into account. We adopt the stability version of finite thickness from \cite{Romeo2011}, which is a combination of gaseous and stellar stability. In most of the calculations, we follow \cite{Zheng2013}. The gas velocity dispersion and surface mass density are obtained from our high-resolution data cube, and the stellar mass density is calculated using WISE-1 band data by using the stellar mass-to-light (M/L) ratio from \cite{Lucero2015}. Notably, the radial component of stellar velocity dispersion is derived from the scale length \citep[1.66$~\mathrm{kpc}$ at a distance of 3.94Mpc;][]{Forbes1992} of the stellar disc, which is described in equation (6) of \cite{Zheng2013}. The total gas disc is assumed to be composed of molecular and neutral gas. The molecular-to-neutral ratio is derived from the SR relation \citep[equation (13) of][]{Zheng2013}, which is an empirical relation between it and stellar surface mass density. The epicyclic frequency is obtained by fitting the universal rotation curve \citep{Persic1996} to our high-resolution RC.

The final disc stability map is shown in the middle panel of Fig.~\ref{fig:sfr_q_fluxratio}. Most of the disc is in an unstable mode with Q values less than 1, which spatially coincides with the SFR surface density (overlaid contours). This suggests that the active star formation in the disc results from gravitational instabilities. Meanwhile, the outskirts of the disc are dynamically stable, which is consistent with former studies that the disc outskirts are stable and the star formation there is suppressed \citep{Meurer2013, Zheng2013}. 

To intuitively understand the correlation between different parameters in a context of asymmetry, we present the radial profile of six properties in Fig.~\ref{fig:radial_profile_6panels}: disc stability, SFR surface density ($\Sigma_{SFR}$), FUV/NIR ratio, \HI\ velocity dispersion, H$\alpha$/FUV ratio and $\SI{70}{\micro\metre}$/$\SI{100}{\micro\metre}$ ratio. The northeast half (approaching side) of the disc is plotted in blue, and the southwest half (receding side) is shown in red. The difference between them is plotted at the bottom of each panel. Significant asymmetries are observed in all distributions. To properly describe them, the disc (3--16$~\mathrm{kpc}$) is divided into 3 parts (their boundaries are marked using green vertical lines in Fig.~\ref{fig:radial_profile_6panels}):

\begin{enumerate}
    \item 3$~\mathrm{kpc}$ < r < 7.5$~\mathrm{kpc}$: Very few asymmetries can be seen in this part. Both sides of the disc are unstable (Q $\sim$0.7) with very large $\Sigma_{SFR}$ (greater than 0.02 \msun\ yr$^{-1}$ kpc$^{-2}$). Although absolute differences of $\Sigma_{SFR}$ between the two sides of the disc are observed (ranging from 0.001 to 0.1 \msun\ yr$^{-1}$ kpc$^{-2}$), they are relatively small proportions to the local $\Sigma_{SFR}$. A nearly constant H$\alpha$/FUV value ($\sim$0.0064) indicates no strong differences in recent SFH between the two sides, which is supported by the FUV/NIR ratio difference plot (bottom of the panel (3)). The FUV/NIR difference is relatively small ($\sim$0.1) compared with the absolute value (2.5-4). $\Sigma_{SFR}$ declines with increasing radius, as well as the dust temperature ($\SI{70}{\micro\metre}$/$\SI{100}{\micro\metre}$) and stellar age estimator (FUV/NIR), indicating a strong correlation between them as formerly discussed. Also, no strong difference in the $\sigma_{\HI\ }$ profiles of the NE and SW could be found. The velocity dispersion decrease with radius on both sides at a similar rate. 
    
    \item 7.5$~\mathrm{kpc}$ < r < 13.75$~\mathrm{kpc}$: In this radius range, the asymmetries are elevated in the distribution of all properties. Firstly, although there is a decreasing trend of $\Sigma_{SFR}$ on both sides of the disc, the $\Sigma_{SFR}$ in the SW disc drops quicker than in the NE. The ratio between $\Sigma_{SFR}$ in NE and SW sides (as shown in the orange line in panel 2) is boosted within this radius range and reaches a maximum of 2.3 at $\sim$11$~\mathrm{kpc}$, which shows that the star formation is far more active in the NE than that in the SW. The H$\alpha$/FUV ratio in NE is also significantly larger than that in SW, as well as the FUV/NIR and $\SI{70}{\micro\metre}$/$\SI{100}{\micro\metre}$ ratios. The differences of all four values ($(\Sigma_{SFR})_{NE}$/${\Sigma_{SFR}}_{SW}$, (H$\alpha$/FUV)$_{NE}$ - (H$\alpha$/FUV)$_{SW}$, (FUV/NIR)$_{NE}$ - (FUV/NIR)$_{SW}$ and $(\SI{70}{\micro\metre}$/$\SI{100}{\micro\metre})_{NE}$ - $(\SI{70}{\micro\metre}$/$\SI{100}{\micro\metre})_{SW}$) share the same trend: They all increase with radius and peak at nearly the same radius of 11$~\mathrm{kpc}$, beyond which the differences all decrease to $\sim$ 0.  It implies that there is excessive star formation activity in the NE, which mainly consists of recent star formation (traced by H$\alpha$) as traced by a younger stellar mean age and higher dust temperature. 
    
    Meanwhile, the two sides of the disc stabilize at different rates within this radius range. At 7.5--11$~\mathrm{kpc}$, the SW side is more efficient in disc stabilization. The Q difference reaches a minimum of -0.1 at $\sim$ 11$~\mathrm{kpc}$, which suggests that the NE disc is less stable at this radius. Then the Q parameter in the NE is quickly boosted compared to the SW, which is probably caused by the feedback of the excessive star formation on the NE side. This explanation is supported by the $\sigma_{\HI\ }$ profile. In contrast to the smaller radius range, the decline of $\sigma_{\HI\ }$ in NE suddenly slows down to zero before $\sigma_{\HI\ ,NE}$ starts to increase again at radii greater than 11$~\mathrm{kpc}$. 
    
    Thus, the excessive star formation intensity on the NE side may be caused by the less stable disc (r < 11$~\mathrm{kpc}$). This might result in higher gas velocity dispersion through star formation feedback, which in turn could then boost the disc stability and suppress the excessive star formation activity at a larger radius (> 11$~\mathrm{kpc}$).
    
    \item 13.75$~\mathrm{kpc}$ < r < 16$~\mathrm{kpc}$: In this radius range, the asymmetries of $\Sigma_{SFR}$, H$\alpha$/FUV and $\SI{70}{\micro\metre}$/$\SI{100}{\micro\metre}$ almost vanish, indicating that the star formation activity is quenched by disc stabilization. The enhanced H$\alpha$/FUV ratio on both sides of the disc probably results from the gas outflow since the dust temperature does not increase with a larger H$\alpha$/FUV ratio as it does at smaller radii. Similarly, the increasing FUV/NIR ratio is no longer a good age estimator since a large portion of the FUV emission stems from the scattering of extraplanar dust, which is expelled from the disc by outflow. Meanwhile, the difference between the Q parameter in the NE and SW is still large, as well as the \HI\ velocity dispersion, which is another consequence of excessive outflow on the NE side. 
\end{enumerate}

Former studies \citep{Davidge2010, Davidge2021} provide evidence, such as an over-luminous northern arm, voids, and bubble-like structures, which implies recent interaction between NGC~253 and its nearby companion NGC~247. This might also have caused the lop-sided morphology of NGC~247 and could explain the recently enhanced SFR on the NE side of NGC~253. Meanwhile, recent observations \citep[eg.][]{Martinez-Delgado2021, Karachentsev2021, Mutlu-Pakdil2022} discovered several nearby dwarf galaxies belonging to the NGC~253 group. The top panel of Fig.~ \ref{fig:nearby_dwarf_anmls_hafuv} shows all confirmed group members within 10 degrees. Their K$_{s}$ band magnitudes, which are gathered from the Local Volume database \footnote{\url{https://www.sao.ru/lv/lvgdb}} \citep{Kaisina2012}, are shown on the top of the plot. They could be seen as a mass proxy since mass measurements for some dwarf galaxies are not available. The 3d distances of the group members to NGC~253 are measured by \cite{Martinez-Delgado2021} using the tip of the red giant branch (TRGB). Three of them (DoIII, DoIV, and SculptorSR) have no distance measurements, and their projected distance is adopted. There are 13 potential members within 700$~\mathrm{kpc}$ distance, which is the zero-velocity radius of the NGC~253 group measured by \cite{Karachentsev2003} by studying the Hubble flow around NGC~253. The crowded environment of NGC~253 could have resulted in an increased rate of interactions with its companions, which is the most likely reason for the asymmetrically enhanced star formation of NGC~253. Notably, most of the NGC~253 group members are located in the north of NGC~253, which coincides with the lopsided distribution of star formation activity. This also implies that the star formation on the NE side of NGC~253 is triggered by interaction.

We conclude that the star-forming disc (3--16$~\mathrm{kpc}$) could be divided into three parts. Intensive star formation is symmetrically distributed in the inner disc (3--7.5$~\mathrm{kpc}$) due to the gravitational instability there (Q$\sim$0.75). At intermediate radii (7.5--13.75$~\mathrm{kpc}$), the star formation on the approaching side (NE) of the disc is possibly triggered by galaxy-galaxy interactions, while the radial profile of the SFR surface density on the receding side is decreasing with radius, similar to other spiral galaxies, which causes the asymmetric patterns of all six properties presented in Fig.~\ref{fig:radial_profile_6panels}. Both the H$\alpha$/FUV and $\SI{70}{\micro\metre}$/$\SI{100}{\micro\metre}$ ratio indicates that the SFR excess has a recent history. An abnormal fluctuation of the Q parameter difference is observed: The NE side is less stable than the SW side at 10--11.5$~\mathrm{kpc}$ and the Q parameter is significantly boosted in the NE side. This is probably a consequence of feedback of the excessive star formation there, which is supported by the \HI\ velocity dispersion profile. At the outskirts of the disc (13.75--16$~\mathrm{kpc}$), asymmetric features disappear in the star formation distribution. Little star formation occurs since the disc is already stabilized at this radius. Meanwhile, the asymmetries in the Q parameter and $\sigma_{\HI\ }$ profile survive, implying the existence of gas outflow on the NE side. This is probably caused by the recently triggered star formation on the NE side in the intermediate radius range (7.5--13.75$~\mathrm{kpc}$).

\subsection{non-declining rotation curve}
\label{subsection:declining_rc}

\begin{figure}
	\includegraphics[width=\columnwidth]{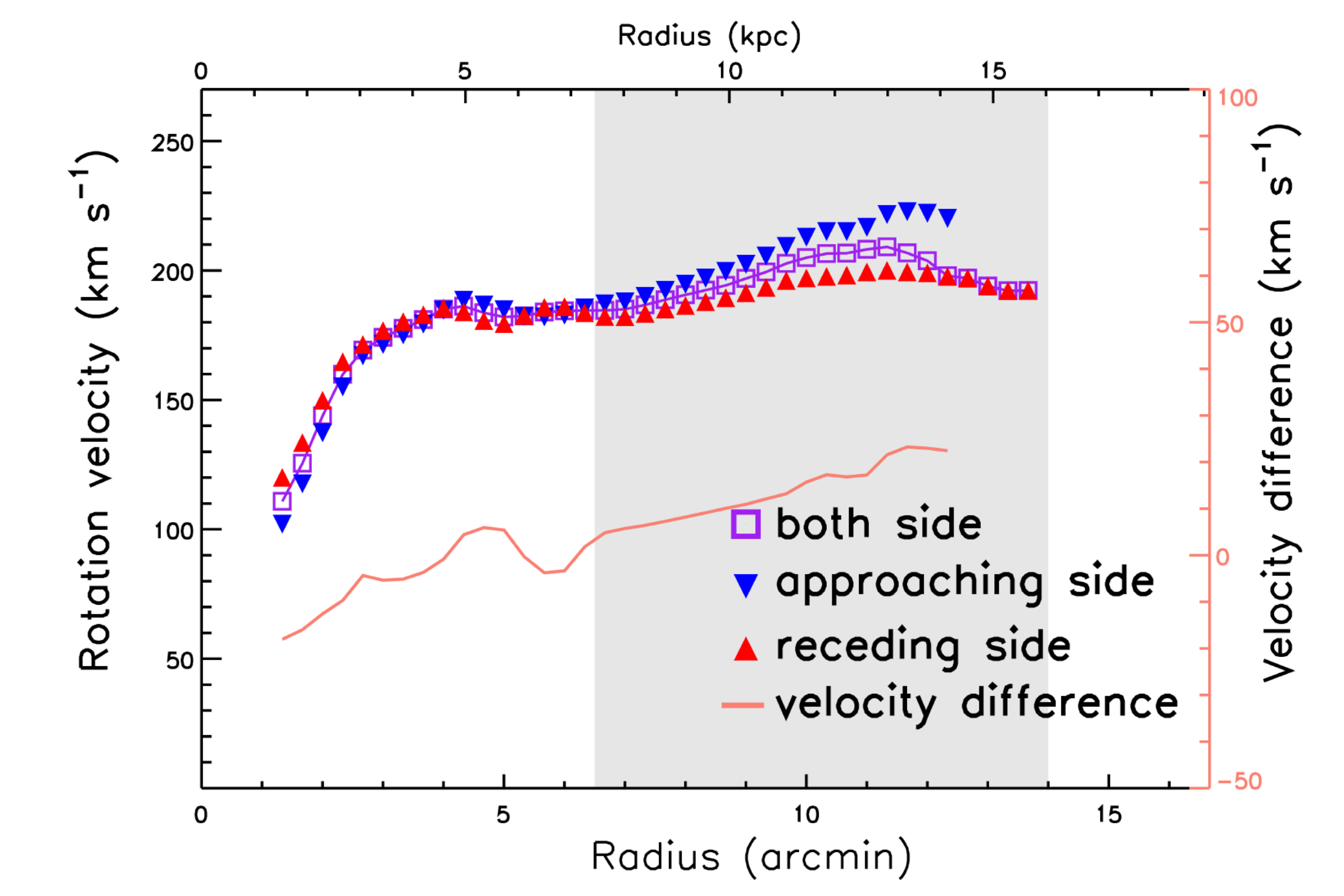}
    \caption{Rotation curve derived from approaching side (blue downward triangle), receding side (red upward triangle), and both sides (purple square). The velocity difference between the approaching side and the receding side is plotted in the orange line using the annotation on the right side. The gray background shows the radius range from $6.5\arcmin$ to $14\arcmin$ (7.5--16$~\mathrm{kpc}$).}
    \label{fig:RCs_both_sides}
\end{figure}

Our rotation curve, which is derived from tilted-ring analysis of both sides of the disc, agrees well with former studies, as shown in Fig.~\ref{fig:RCs_pure}. Furthermore, we obtained a high-resolution RC out to 14 \arcmin\ ($\sim$16$~\mathrm{kpc}$), 2 \arcmin\ beyond the transition radius (12 \arcmin;\ 13.75$~\mathrm{kpc}$), where the rotation velocity starts to decrease according to \cite{Bland-Hawthorn1997} and \cite{Hlavacek-Larrondo2011}. Our RC from both sides of the disc successfully reproduces this trend. The rotation velocity reduces by $\sim$ 20 \kms\ compared to the highest point (209.2 \kms\ ) at the outskirts of the disc. Thus, the rotation curve of the \HI\ disc seems to be declining.

\begin{figure}
	\includegraphics[width=\columnwidth]{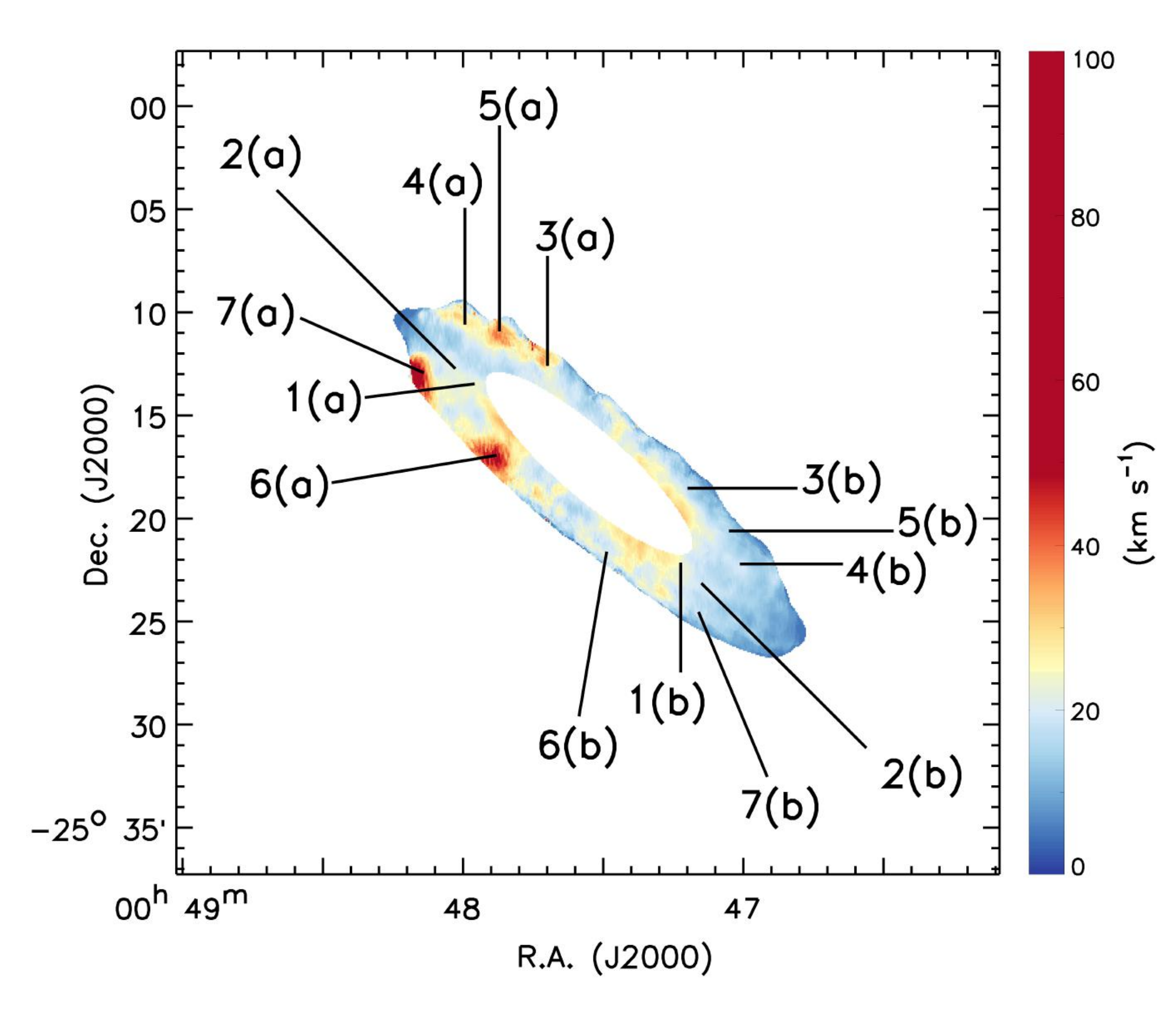}
    \caption{Moment 2 map of radius from 6.5 to 14 arcmin. Several pixels in the interesting location were selected. They were labelled in numbers and letters. With a larger labelled number, the radius of the labelled pixel is also larger. The pixels with the same number but different letter labels are symmetric about the minor axis.}
    \label{fig:moment2_outer}
\end{figure}

\begin{figure}
	\includegraphics[width=\columnwidth]{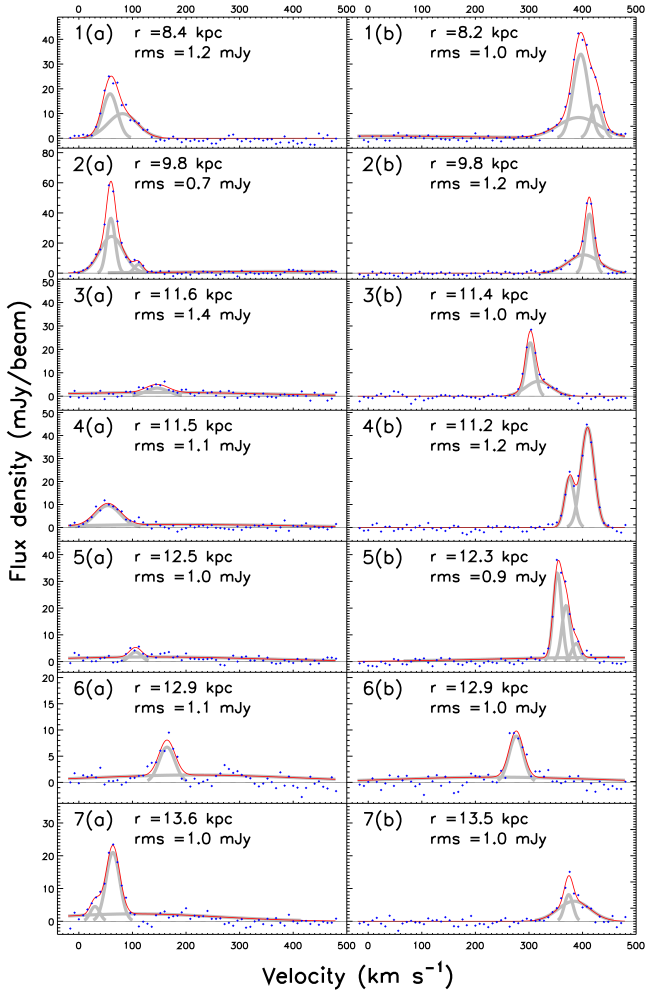}
    \caption{Line profiles of the \HI\ emission of the positions labeled in Fig.~\ref{fig:moment2_outer}. The recognized components are shown in grey lines. The final fitted results are plotted in red lines. The observed data points are shown in blue dots. The radius and noise level of each pixel is also shown in the plot.}
    \label{fig:pixtrum_plot}
\end{figure}

However, further inspection of the RC on each side of the disc reveals a different answer. Both the approaching side and the receding side (Fig.~\ref{fig:RCs_both_sides}) have a flat RC on the outer disc ($10\arcmin$<R), which is in contrast to the declining trend of the RC from both sides (hereafter RC$_{both}$). Additionally, the errors of the RC$_{both}$, normally derived from the velocity difference of the receding and approaching side, are suspiciously large at the transition radius, which implies an asymmetric rotation. In fact, the large uncertainties of RC$_{both}$ in this radius range are also observed in all previous studies \citep{Puche1991, Lucero2015, Hlavacek-Larrondo2011}. The asymmetric features of the rotation curve can be understood in two ways:

\begin{enumerate}
\item The receding side extends further ($\sim 2\arcmin$) than the approaching side. Fewer data points from the approaching side contribute to the RC fit at radii of $12\arcmin < R \leqslant 14\arcmin$, where the previously reported declining RC is observed \citep[also see section 6.1.2 in ][]{Lucero2015}.
\item The rotation velocity on the approaching side (maximum velocity $\sim$220\kms\ ) is significantly higher than on the receding side (maximum velocity $\sim$190\kms\ ) in the outer disc. The velocity difference increases with radius up to $\sim$30 \kms.\

\end{enumerate}

This suggests that the reducing rotation velocity of both sides in the outer disc is the combined effect of these asymmetries: the measured rotation velocity is boosted up by the approaching side at $6.5\arcmin$(7.5kpc) < R $\leqslant 12\arcmin$(13.75$~\mathrm{kpc}$) while the total RC drops at R > $12\arcmin$(13.75 $~\mathrm{kpc}$) where it is dominated by the receding side, the rotation velocity of which is much lower. This explanation is not in conflict with former observations since similar features can be noticed in the RC results from all previous studies that reported a declining RC. However, those studies simply draw conclusions from the combined RC, which is strongly affected by the significant kinematical asymmetries. 
Considering the presence of strong asymmetries in the outer disc, the combined RC is no longer a precise tracer of the underlying mass distribution. Consequently, the declining trend of the RC found in previous studies does not necessarily imply a truncation of the dark matter halo. The NGC~253 case suggests that the conclusions of a declining rotation curve should be reached with caution, especially for those galaxies that are kinematically asymmetric at large radii. In fact, signs of similar asymmetries can be seen in several recent studies perceiving a declining rotation curve (Figure~2 in \cite{Casertano1991}, Figure~5 in \cite{Dicaire2008}).

In addition to the increasing difference of rotation velocities, the velocity dispersion (panel 4 of Fig.~\ref{fig:radial_profile_6panels}) on the approaching side (NE) is also systematically higher than on the receding side (SW). The difference between velocity dispersion profiles of two sides of the disc grows with radius up to $\sim$25 \kms\ at $6.5\arcmin < R \leqslant 14\arcmin$, which coincides with the radius range of asymmetric RCs. This implies that the asymmetric features are correlated with, and probably caused by the turbulence of the \HI\ disc on the NE side. 

To get a detailed insight into the turbulence at $6.5\arcmin < R \leqslant 14\arcmin$, the moment-2 map of this radius range is shown in Fig.~\ref{fig:moment2_outer}. Several turbulent features are clearly visible on the approaching side. To explore the \HI\ emission of these features, a few typical pixels were chosen (labelled in Fig.\ref{fig:moment2_outer}). The \HI\ emissions and their Gaussian-decomposition results are plotted in Fig.~\ref{fig:pixtrum_plot}. These illustrative examples suggest a growing dominance of the anomalous gas over the \HI\ emission with increasing radius on the approaching side (NE), which reflects the turbulent nature there. In addition, the \HI\ line profile at these positions in the low-resolution data cube is also presented in a similar figure, which is available in the supplementary online material. The anomalous gas in low-resolution line profiles is more significant, which suggests that the \HI\ disc is truly disrupted in these regions.

Moreover, the turbulence of the \HI\ disc is spatially correlated with the elevated star formation. Both occur at the same radius range (6.5--14$\arcmin$; 7.5--13.75$~\mathrm{kpc}$), which is emphasized in Fig.~\ref{fig:radial_profile_6panels} via the grey background. As discussed in section~\ref{subsection:star_forming_disc}, the extra turbulent gas on the NE side is probably the gas outflow triggered by the excessive star formation, especially that traced by H$\alpha$ emission, at the intermediate disc radii.

Therefore, instead of ending up with the conclusion of a truly declining rotation curve, we find that the rotation curve is truly asymmetric. The declining features observed by former studies are actually caused by the asymmetries of the \HI\ kinematics, which result from asymmetric star formation feedback. Excessive star formation activities on the approaching side deposit energy into the \HI\ gas, which eventually could result in asymmetric rotation curves on each side of the disc. This assumption could be examined by checking whether the extra-planner gas is caused by outflow or not, which will be discussed in section~\ref{subsection:anomalous_gas}.

\subsection{anomalous gas}
\label{subsection:anomalous_gas}

\begin{figure}
	\includegraphics[width=\columnwidth]{nearby_dwarf_anmls_column_density_ha_fuv_cvlv_blue_pink.pdf}
    \caption{Top: confirmed members of the NGC~253 group within 10 degrees radius. Only 6 of them have velocity measurements, which are indicated by the red numbers (velocities relative to NGC~253) below the symbols. All galaxies are colourized according to their 3d distance to NGC~253 (NGC~253 itself is shown as the black star). Three group members (DoIII, DoIV, and SculptorSR), which lack distance measurements, are labelled with dashed blue squares. Their colours are chosen according to their projected separation from NGC~253. NGC~247 is emphasized as it is by far the most massive companion of NGC~253. All galaxies are symbolized according to their K$_{s}$ band luminosity, as shown at the top of this figure. Bottom: Combined \HI\ moment-0 map (blue), stellar mass surface density (yellow contours) and \Ha\ to FUV flux ratio (purple). The anomalous gas is plotted in white contours, which correspond to column densities of 4, 12 and 40$\times$ $10^{18}$~cm$^{-2}$. Regions with higher column densities are indicated by thicker contours.}
    \label{fig:nearby_dwarf_anmls_hafuv}
\end{figure}

\begin{table}
	\centering
	\caption{\HI\ mass and anomalous \HI\ mass from this work compared with former studies.}

	\begin{tabular}{ p{1.6cm}llc  } 
		\hline
		     & M$_{ \HI\ }$ (anomalous) & M$_{\HI\ }$ (total) & mass ratio \\
		\hline
		this work & 1.02$\pm{0.12}$ $\times$ $10^{8}$ \msun\ & 2.49$\pm{0.023}$ $\times$ $10^{9}$ \msun\ & 0.041\\

		Boomsma et al. (2005) & 8 $\times$ 10$^{7}$ \msun\ & 2.5 $\times$ 10$^{9}$ \msun\ & 0.03\\

		Lucero et al. (2015) & 7.8 $\times$ 10$^{7}$ \msun\ & 2.1 $\times$ 10$^{9}$ \msun\ & 0.035\\
		\hline
	\end{tabular}
	\small
    Notes. Anomalous and total refer to anomalous \HI\  mass and total \HI\ mass found in the low-resolution data cube, respectively. The corresponding statistical uncertainties are calculated using the RMS noise of the data cube. The mass ratio is the ratio between the two.
    \label{tab:hi_anmls_mass}
\end{table}

As introduced in section~\ref{subsection:anmls_subtraction}, we isolate the anomalous \HI\ via the combination of Gaussian decomposition results and the modelled velocity map from RC fitting, which is different from the separation method adopted by former studies (visual inspection). A comparison of the \HI\ mass of anomalous gas and total gas between different studies is available in Table.~\ref{tab:hi_anmls_mass}, the corresponding statistical uncertainties of which are calculated using the RMS noise of the final data cube. 
Notably, there is an obvious difference in the total \HI\ mass measurements between our data (2.5 $\times 10^{9}$ \msun)\ and that of \cite{Lucero2015} (2.1 $\times 10^{9}$ \msun).
This is not caused by the statistical uncertainties of the flux measurements but mainly a consequence of the significant HI absorption feature against the bright star-forming inner region of NGC~253, which is visible as the central hole in Fig.~\ref{fig:moments} and Fig.~\ref{fig:nearby_dwarf_anmls_hafuv}. Our data have a higher angular resolution ($154\times81$ \arcsec versus $213\times188$ \arcsec), which means that less flux is removed by the \HI\  absorption through beam smearing. This trend is also seen in the mass measurements in \cite{Boomsma2005} (2.5 $\times 10^{9}$ \msun; 70 \arcsec) and \cite{Koribalski2018} (2.7 $\times 10^{9}$ \msun; $30\times70$ \arcsec). Considering the existence of the \HI absorption, all the \HI\ mass measurements should be treated as  lower limits as flux is inevitably lost due to the negative signal in the centre. However, it does not seriously affect the mass measurement of the anomalous \HI\ which is mainly located in the outer disc.

We find anomalous \HI\ mass (1.02 $\pm{0.12}$ $\times$ $10^{8}$ \msun\  corresponding to 4.1 percent of total \HI\ mass) is slightly larger than former studies. This is probably caused by the better ability of our toolkit to isolate non-rotational gas components in complex line profiles. The bottom panel of Fig.~\ref{fig:nearby_dwarf_anmls_hafuv} shows the spatial distribution of anomalous \HI.\ Notably, the anomalous \HI\ within the disc is difficult to be decomposed since it overlaps with the disc emission, which is much stronger. As a result, only the strongest ones are recognized, which results in the discrete distribution of the anomalous gas within the disc region. To get a smoother and more illustrative view of its spatial distribution, we convolve both the \HI\ moment 0 map and the subtracted anomalous gas map with a $1\arcmin$ Gaussian PSF. The convolution has almost no impact on the moment 0 map given that the original beam is much larger. Fig.~\ref{fig:nearby_dwarf_anmls_hafuv} shows that most of the anomalous gas is located on the NE side of the disc, where chimney-like structures vertically extend out of the disc in both directions. On the SW side, little anomalous gas is found, but its morphology is similar to that in the NE, albeit less extended. Additionally, the anomalous gas on the NE side, especially that with column density higher than 4$\times$ $10^{19}$~cm$^{-2}$, spatially correlates with higher H$\alpha$/FUV flux ratios, which implies a correlation between anomalous gas and recent massive stellar feedback.

\begin{figure}
	\includegraphics[width=\columnwidth]{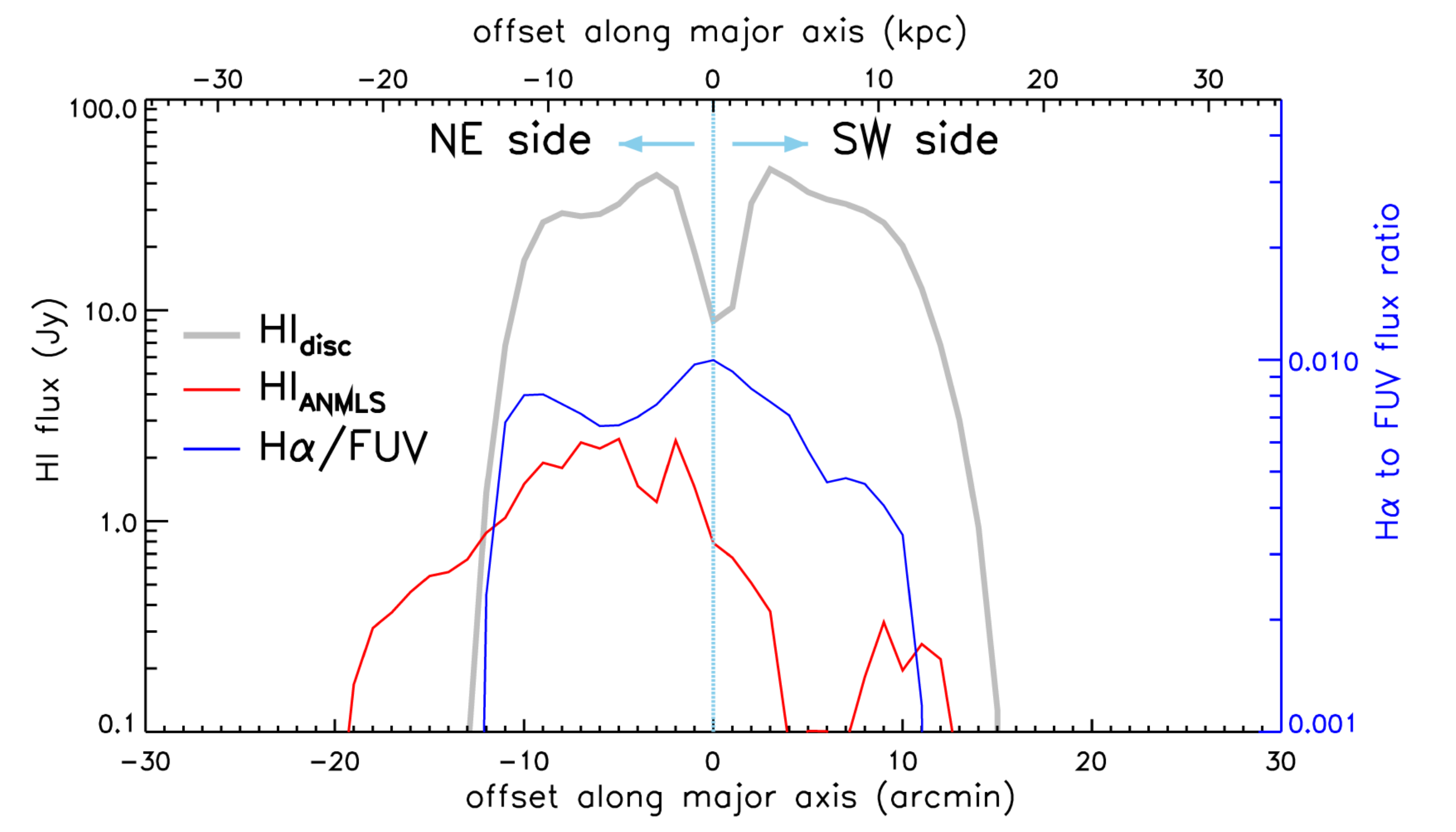}
    \caption{\HI\ flux density profile versus angular offset along the major axis. The flux density is calculated by summing the pixels along the direction of the minor axis. The anomalous \HI\ is plotted in red, and the remaining \HI\ gas in the rotating disc is shown in thick grey lines. The cumulative H$\alpha$/FUV flux ratio is also overplotted in blue.}
    \label{fig:hi_dist_profile}
\end{figure}

\begin{table}
	\centering
	\caption{The mass of anomalous \HI\ and disc \HI\ for both halves of the disc.}
	\begin{tabular}{ p{2cm}llc  } 
		\hline
		     &  NE (approaching)  & SW (receding) & ratio\\
		\hline
		M$_{ \HI\ }$(anomalous) & 8.89 $\times$ $10^{7}$ \msun\ & 1.31 $\times$ $10^{7}$ \msun\ & 6.78$\pm{1.23}$ \\

		M$_{ \HI\ }$ (disc) & 1.15 $\times$ $10^{9}$ \msun\ & 1.24 $\times$ $10^{9}$ \msun\ & 0.93$\pm{0.0056}$ \\

		M$_{ \HI\ }$ (total) & 1.24 $\times$ $10^{9}$ \msun\ & 1.25 $\times$ $10^{9}$ \msun\ & 0.99$\pm{0.014}$ \\
		\hline
	\end{tabular}
	\small
     Notes. Similarly, anomalous, disc and total refer to anomalous \HI,\  disc \HI,\  and the sum of them, respectively. The ratio refers to the mass ratio between the NE and SW side of the disc.
    \label{tab:hi_mass_NW_SE}
\end{table}

Since the anomalous gas is unevenly distributed on the two sides of the disc, it is more instructive to show the cumulative \HI\ profile versus the offset along the major axis with negative offsets referring to the NE side of the disc. Fig.~\ref{fig:hi_dist_profile} shows the cumulative flux density profile of both rotational (disc \HI\ ) and non-rotational (anomalous \HI\ ). As expected, extensive and excessive anomalous gas can be seen in the NE. Also, the receding (SW) half of the \HI\ disc extends $\sim$2 arcmin further than the approaching (NE) half. The \HI\ mass of anomalous and disc components on the NE and SW sides and the mass ratio between the two sides are summarised in Table~\ref{tab:hi_mass_NW_SE}. The corresponding uncertainties of the mass ratios are calculated via the propagation of flux measurement uncertainties. The mass ratio shows a clear trend: there is far more ($\sim$7 times) anomalous gas on the NE side than in the SW. However, the total \HI\ on both sides of the disc is nearly equivalent. This trend is still significant after the uncertainties of the flux measurements have been taken into account.

It is therefore reasonable to assume that most of the anomalous gas on the NE side, which caused the uneven distribution of the anomalous gas across the two halves of the disc, is former disc gas that was expelled by processes associated with the excessive star formation there. In principle, gas inflow from some unknown source could potentially have added anomalous gas preferentially to the NE side of NGC 253, although we consider such a scenario unlikely given the evidence for outflow presented in this work.

\cite{Davidge2021} provides clues to a possible interaction between NGC~247 and NGC~253, which may have caused the lopsided mophology of NGC~247. In addition to the overluminous northern spiral arm, \cite{Davidge2021} also identified two kpc-size bubbles in the disc of NGC~247 using the deprojected UV and IR images. Under the assumption that these bubbles are the shells of ISM expansion caused by star-forming activity and by applying a constant expansion speed of 7 \kms\ (which is the expansion velocity of similar structures in the disk of the dwarf galaxy Holmberg II measured by Puche et al. (1992)), he derives dynamical ages of 230 and 150 Myr for the south and north bubbles, respectively. These ages agree with the age range (100--300 Myr) of the recent SFR enhancement in the nuclear and circumnuclear regions of NGC~247 \citep{Kacharov2018}, which supports the assumption that these bubbles originate from ISM expansion triggered by stellar feedback. Meanwhile, \cite{Davidge2010} suggests that the enhanced star formation on the NE side of NGC~253 occurred within at least the past a few tens of Myr and probably results from the interaction with NGC~247. Since the extraplanar \HI\ of NGC~253 is another possible consequence of this interaction, it is also worthwhile for us to estimate the kinematical age of the extraplanar gas and compare it with the former values.

\begin{figure}
	\includegraphics[width=\columnwidth]{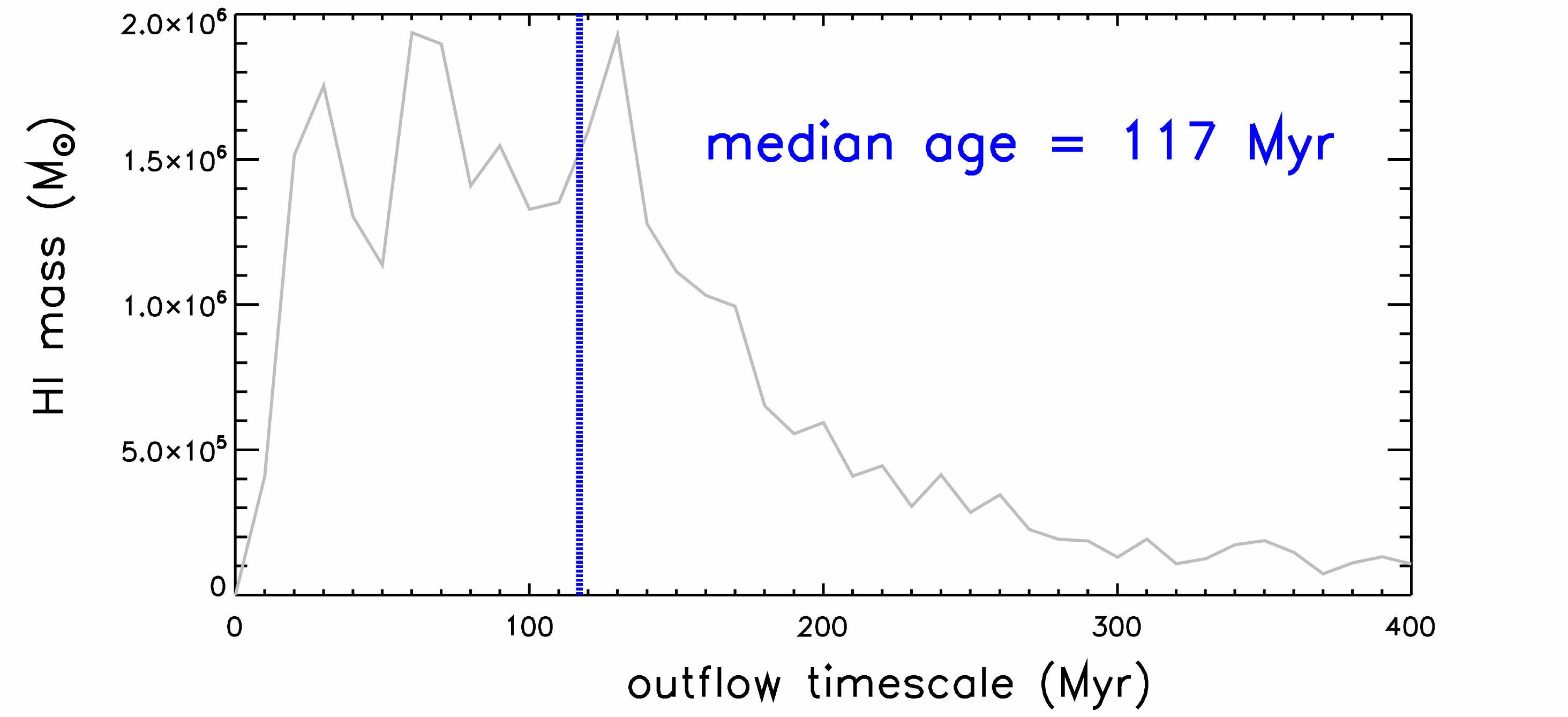}
    \caption{Estimated outflow time for extraplanar anomalous \HI\ versus \HI\ mass in each time bin (10 Myr). The median timescale is shown as the vertical blue dashed line at 117 Myr. Considering the inclination of NGC~253($\sim$$\ang{78}$), this value is a rough estimation.}
    \label{fig:outflow_timescale}
\end{figure}

By visual inspection of the moment-0 map of the low-resolution data, we adopt 19~kpc as the boundary of the \HI\ disc. Therefore, we define the anomalous gas at R >19kpc as the extra-planar gas. All pixels containing the extra-planar gas are selected. For each pixel, we adopt the central velocity from our Gaussian-decomposition fitting as the line-of-sight velocity of the extra-planar gas. Then, the velocity difference is calculated by comparing the central velocity with the modelled velocity from RC fitting. Finally, the timescale of the anomalous outflow in each pixel is obtained by dividing the absolute value of velocity difference by the projected distance to the major axis. Fig.~\ref{fig:outflow_timescale} shows the timescale distribution versus the mass in each timescale bin (10 Myr). A timescale range of 0-800 Myr is selected as it covers more than 90 percent of the extraplanar gas. The median outflow age is approximately 120 Myr. Notably, considering the high inclination of NGC~253, the timescale estimation is very rough due to projection effect. The median timescale is approximately consistent with the timescale ranges indicated by \cite{Davidge2010} (at least a few tens of Myr) and \cite{Davidge2021} (150-230 Myr), which suggests that the outflow of extra-planar \HI\ of NGC~253 is likely correlated with the possible interaction with NGC~247. Furthermore, one disk rotation time is $\sim$360 Myr at the radius of 11$~\mathrm{kpc}$, where the asymmetry of the star formation reaches its maximum. This also implies that the extraplanar gas is caused by outflow since the timescale of inflow or accretion should be at least several times the disc rotation for gas to lose angular momentum. 

Therefore, based on the mass distribution analysis and timescales estimation, the extended \HI\ structures, especially those on the NE side, are likely to be gas outflow caused by enhanced star formation triggered by galaxy-galaxy interaction. Although \cite{Davidge2010} also mentioned a causal relationship between the elevated SFRs on the NE side and the extra-planar \HI\ chimneys there, there is another issue left: is the star formation on the NE side of the disc active enough to support the chimney-like outflow? Through high-resolution ($\sim$35 pc) simulations of stellar feedback (mainly supernova (SN) explosions and stellar winds) on interstellar medium (ISM), \cite{Ceverino2009} successfully reproduce kpc-scale galactic chimneys and find a nearly constant SFR surface density after an initial burst (around 0.005 \msun\ $yr^{-1}$ $kpc^{-2}$; shown in the top panel of Figure 3 in \cite{Ceverino2009}). This condition could be easily satisfied in the case of NGC~253 since most of the regions on the NE side have a SFR surface density similar to 0.01 \msun\ $yr^{-1}$ $kpc^{-2}$ (shown in the panel (2) of Fig.~\ref{fig:radial_profile_6panels}) at 7.5<r<13.75$~\mathrm{kpc}$, where asymmetries in the star formation activity are clearly observed as discussed in section.~\ref{subsubsection:sfr_and_other_properties}. The top panel of Fig.~\ref{fig:sfr_q_fluxratio} also shows several \HII\ regions with SFR surface densities > 0.02 \msun\ $yr^{-1}$ $kpc^{-2}$ in the NE spiral arm (they are also highlighted by the contours of \Ha\ to FUV flux ratio). This means that the NE side of NGC~253 is potentially powerful enough to form the chimney structures.

\section{SUMMARY AND CONCLUSIONS}
\label{section:conclusions}

In this paper, we present a multi-wavelength study of star formation feedback on the kinematics of the ISM in a specific galaxy: NGC~253. The three well-known features of NGC~253 (a disrupted stellar disc, a previously reported declining rotation curve, and the anomalous \HI\ gas) are studied in detail in a common context of asymmetry and are found to be connected by the mechanism of star formation feedback. Our main results and conclusions are summarized below.

\begin{enumerate}
    \item We gathered all ATCA \HI\ observations available since the telescope was built. After careful data reduction (in which the role of the multi-scale cleaning technique is emphasized), we create two versions of \HI\ data cubes with high angular resolution (30$\arcsec$) and deep column density sensitivity ($\sim$4$\times$ $10^{18}$~cm$^{-2}$) respectively. The high-resolution data provide precise kinematical measurements of \HI\ emission out to 14$\arcmin$ ($\sim$16$~\mathrm{kpc}$) to resolve and trace the RC up to 2\arcmin\ beyond the transition radius (where former studies found the RC start decreasing). Meanwhile, the low-resolution data are perfect for studying anomalous gas due to their high column density sensitivity.

    \item To properly separate the anomalous gas from overlapping the disc gas, we also develop our own toolkit called FMG to perform a Gaussian decomposition of the \HI\ emission line profiles, which automatically separates the spectrum into different kinematical components and estimates their parameters separately. By combining the $\chi ^{2}$ minimization technique and BIC theory, our toolkit provides fast (compared to the MCMC method) and reliable parameter estimation of different kinematical components in complex line profiles (especially those with 2-4 components), which has been tested on mock spectra (section~\ref{subsubsection:mock_spectrum_test}). Most importantly, our toolkit has been proven to be capable of recognizing extremely broad components in complex line profiles with good precision, which makes it a good tool to isolate anomalous gas from the rotating disc.
    
    \item To explore the enhanced star formation activities in NGC~253's disc, both H$\alpha$ and FUV images are analyzed. The foreground stars are identified and removed based on their UV colours and G-band magnitudes (from Gaia). Dust attenuation effects are corrected using the empirical relation proposed by \cite{Kennicutt2009} and \cite{Hao2011}, which linearly combines the H$\alpha$ and FUV emission with the total infrared luminosity. Since this is the first time that these formulas are used for resolved dust attenuation, we also test our procedure on the pixel-by-pixel IRX-$\beta$ relation. We find that NGC~253's disc could be perfectly described by the IRX-$\beta$ relation from \cite{Meurer1999} for starburst galaxies. In addition, a tight correlation is discovered between H$\alpha$ luminosity, H$\alpha$/FUV, $\SI{70}{\micro\metre}$/$\SI{100}{\micro\metre}$ flux ratio and the perpendicular distance to Meurer99 relation. This not only suggests that our dust correction is reliable but also indicates that the recent/massive star formation spatially correlates with dust temperatures. 
    
    \item To detailedly study the asymmetries of the disrupted disc, we investigate the radial profile of 6 correlated properties (stability parameter, \HI\ velocity dispersion, SFR surface density,  H$\alpha$/FUV flux ratio,  stellar age estimator traced by FUV/NIR and dust temperature tracked by $\SI{70}{\micro\metre}$/$\SI{100}{\micro\metre}$) across the two halves of the disc (NE, also approaching side; SW, also receding side). We find that NGC~253's disc can be divided into 3 parts according to their levels of asymmetry. The inner region (3--7.5$~\mathrm{kpc}$) is symmetrically unstable (Q$\sim$0.7) and bursty ($\Sigma_{SFR}$ 0.02-0.12 \msun\ yr$^{-1}$ kpc$^{-2}$). At intermediate disc radii (7.5--13.75$~\mathrm{kpc}$), asymmetric features can be clearly observed on profiles of all six properties (shown as Fig.~\ref{fig:radial_profile_6panels}). Excessive star formation on the NE side, mainly traced by H$\alpha$, is clearly visible. The differences in $\Sigma_{SFR}$ between the two sides grow with radius, peak at 11$~\mathrm{kpc}$, and eventually vanish at 13.75$~\mathrm{kpc}$. This trend is accompanied by differences in stellar age, dust temperature, ISM velocity dispersion, and disc stability between the two halves. All of these parameters follow a similar trend and peak at a similar radius. It suggests that the star formation on the NE side (more preferably in H$\alpha$ sense) is enhanced, which heats the dust and causes gas outflows. The ISM is subsequently disturbed, which stabilizes the disc and, in turn, suppresses the star formation at larger radii. In the outskirts of the disc (13--16$~\mathrm{kpc}$), since both sides of the disc are stable, most asymmetric features disappear except for the Q parameter and \HI\ velocity dispersion, which is affected by outflow. 
    
    \item By fitting a tilted-ring model to the \HI\ velocity field derived from the high-resolution data, we obtain the high-resolution rotation curve out to a radius of 14 \arcmin.\ The RC, fitted using data points from both sides of the disc, reproduces the declining trend of the rotation velocity reported by former studies. However, closer inspection of the RCs from each half of the disc separately reveals that the RCs on both sides of the disc are flat at large radii. The combined RC is not naturally delining but becoming increasingly asymmetric in the outer disc (7.5--16$~\mathrm{kpc}$). The declining trend is the combined effect of two aspects of asymmetry: (1) the combined rotation velocity is boosted up by the SW side at 7.5-13.75 kpc, since the rotation velocity there is systematically higher (up to $\sim$30 \kms\ ) than SW side. (2) the combined rotation velocity quickly drops to the same velocity ($\sim$190 \kms\ ) as that on the SW side at 13-16 kpc since the NE disc does not extend to this radius and the SW side dominates the combined RC fitting. This not only challenges the previous perception that NGC~253 has a declining RC but also provides an instructive clue on the future analysis of similar cases: It is important to take the asymmetries into account when reaching the conclusion of a declining RC.

    Meanwhile, a systematically higher velocity dispersion at 7.5--16$~\mathrm{kpc}$ radius on the NW side reflects the turbulent nature there. A few representative \HI\ emission line profiles of the turbulent features suggest that they are caused by the growing dominance of the anomalous component over the \HI\ disc emission on the NW side with increasing radius. This trend spatially coincides with the asymmetries of the RC, which indicates that the perceived declining trend of the RC is probably caused by the presence of the anomalous gas.
    
    \item We successfully isolate the anomalous gas from the \HI\ disc by their velocity dispersion (anomalous components are intrinsically broader than rotational ones) and the difference between the observed and modelled velocity field from RC fitting. We find a 20\% larger anomalous HI mass of 1.02 $\times$ $10^{8}$ \msun\ as compared to previous studies, which is reasonable since our low-resolution data cube is more sensitive and our toolkit is better capable of separating anomalous components than visual inspection. Meanwhile, the structure of the anomalous gas is very similar to former studies: most anomalous gas is located on the NE side where spurs vertically extended from the disc up to $\sim$12 kpc away. The \HI\ mass distribution suggests that there is 7 times more anomalous \HI\ in the NE than in the SE, which is consistent with the asymmetric distribution of the star formation intensity. This implies the extraplanar anomalous \HI\ was expelled from the disc by star formation feedback. More importantly, the total \HI\ (anomalous + disc) on both sides are perfectly equivalent, which suggests an outflow origin of the extraplanar \HI.\ 
    
    In addition, we also estimated the outflow timescale of extraplanar \HI\. The median outflow timescale from our estimate ($\sim$120 Myr) corresponds to about one-third of one disk rotation, which supports the outflow hypothesis since inflow/accretion timescales are expected to be larger than a few disc rotation periods for gas to lose its angular momentum. Also, our outflow timescale agrees well with the interaction timescale estimations for NGC~253 from previous studies \citep{Davidge2010, Davidge2021}. The spatial distribution of the NGC~253 group (top panel of Fig.~\ref{fig:nearby_dwarf_anmls_hafuv}) indicates a high density of satellite galaxies in the vicinity of NGC 253, which implies a high chance of recent interaction. Such interactions are likely to have triggered the asymmetric star formation and consequently gas outflow. A previous study by \cite{Ceverino2009} has shown that the star formation activity traced by the \HII\ regions in the NE arm is active enough to form the chimney-like outflow of NGC~253.

\end{enumerate}

\section*{Acknowledgements}

XL acknowledges useful suggestions from Brent Groves (on foreground star removal), Luca Cortese (on dust attenuation), Danail Obreschkow (on BIC theory), Matthew Young (on general science), Qingxiang Chen (on CASA usage), Pei Zuo (on general science), Sambit Roychowdhury (on dust attenuation), and Simone Bianchi (on the Herschel-PACS PSF). We thank the anonymous referee for their very useful comments. The Australia Telescope Compact Array is part of the Australia Telescope National Facility (https://ror.org/05qajvd42) which is funded by the Australian Government for operation as a National Facility managed by CSIRO. This paper includes archived data obtained through the Australia Telescope Online Archive (http://atoa.atnf.csiro.au). 

\section*{DATA AVAILABILITY}

The IDL toolkit FMG (Fit Multiple Gaussian components) and the data underlying this article will be shared on reasonable request to the corresponding author.








\appendix
\section{details of the fitting toolkit: FMG}
\label{section:details_toolkit}

\begin{figure}
	\includegraphics[width= \columnwidth]{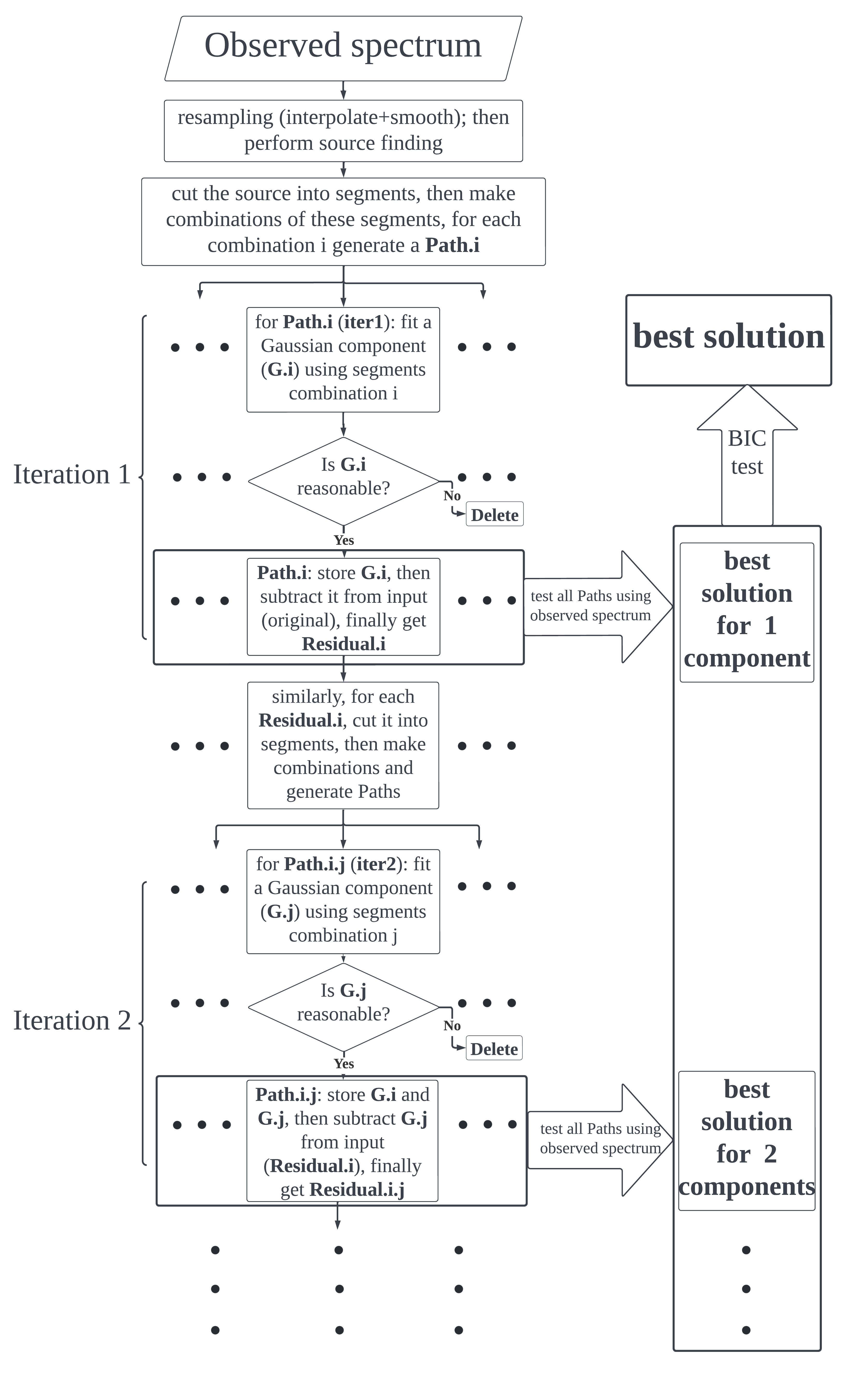}
    \caption{Flow chart of Gaussian decomposition toolkit.}
    \label{fig:fitting_flow}
\end{figure}

Generally, there are two kinds of fitting procedures for multiple Gaussian components. Those based on the $\chi^2\ $ minimization technique \citep{Nidever2008, Lindner2015, Ho2016} converge quickly but rely heavily on initial parameter estimates because they often converge at a local minimum without a proper sampling of the parameter space. Also, they often suffer from overfitting problems due to difficulty in deciding the number of components being used. The Markov Chain Monte Carlo (MCMC) technique \citep{Oh2019} is immune to this problem but requires much more computational resources since it computes the likelihood distribution of the parameter space. Once the fitting is finished, the models with different numbers of components should be compared using statistical model selection criteria, like the Bayesian information criterion \citep[BIC; ][]{Schwarz1978} or the Akaike information criterion \citep[AIC; ][]{Akaike1974}, to prevent over-fitting.

We developed  an IDL toolkit called FMG (Fit Multiple Gaussian components) based on the $\chi^2\ $ minimization procedure to fit multiple Gaussian components to \HI\ data cubes. The local minimization problem was prevented by supplying proper initial guesses, and the final results were assessed using the BIC. We adopt the $\chi^2\ $ minimization technique for two reasons:

\begin{figure}
	\includegraphics[width=\columnwidth]{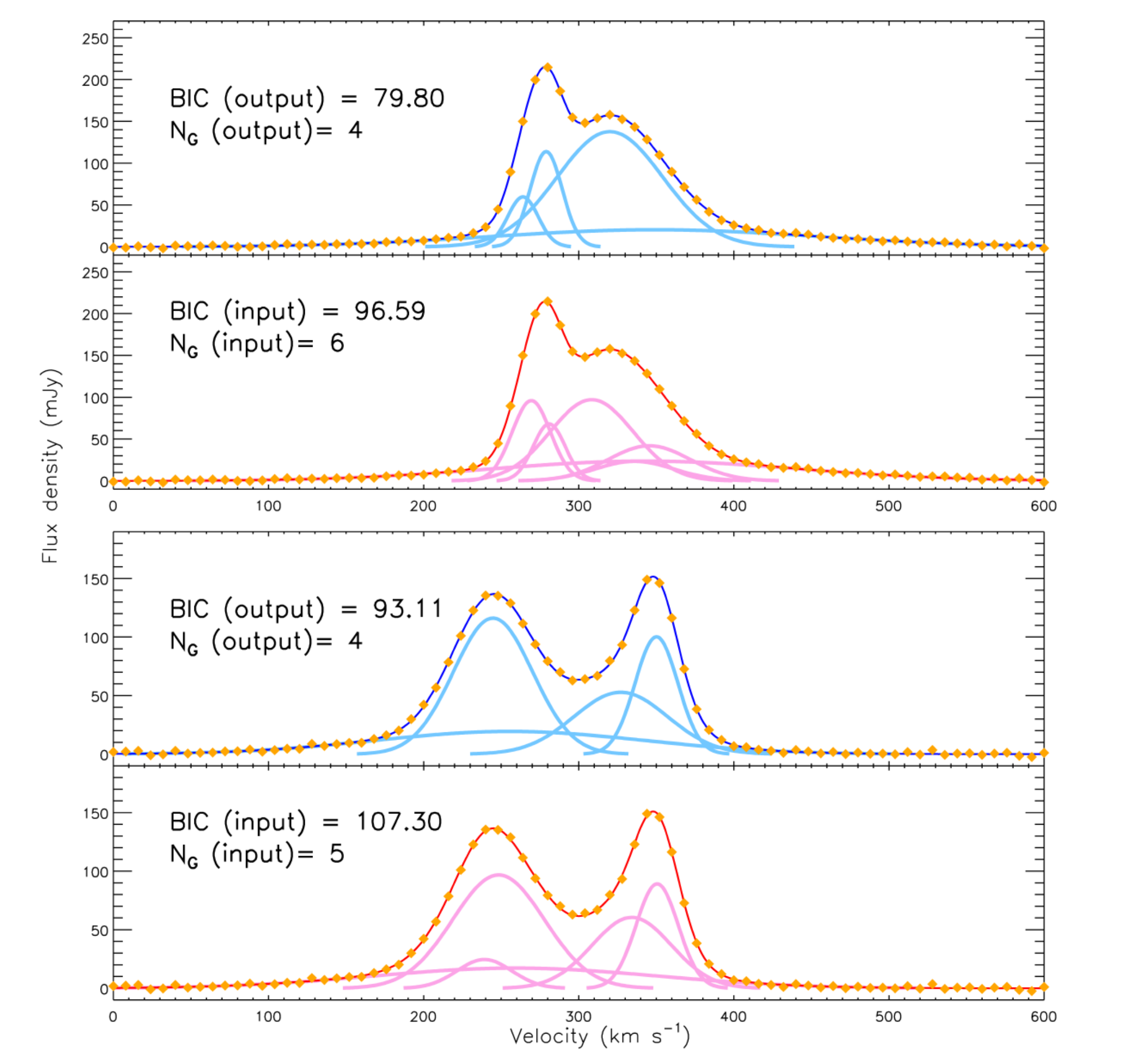}
    \caption{Two examples of ``merged'' Gaussian components in complex line profiles. The modelled data points (Gaussian components + noise) are plotted in orange. Input components are plotted in pink and decomposed results were plotted in blue. The final BIC value and the number of Gaussian components (N$_{G}$) are also shown in each panel.}
    \label{fig:mock_spec_G5_6}
\end{figure}

\begin{enumerate}
\item There are a total of $\sim$60,000 spectra in the high-resolution and low-resolution data cubes to fit and a maximum of 6 possible components (18 free parameters in total) for one single spectrum. The huge computational resources required for the MCMC technique are not available to us.
\item The detailed likelihood distribution for the whole parameter space, which is provided by the MCMC, is sometimes not necessary for complex line profiles (>5 components). Because if too many (>5) Gaussian components overlap with each, they will statistically ``merge'' into fewer components, which makes it impossible to obtain a perfect estimate of the original parameter. For example, Fig.~\ref{fig:mock_spec_G5_6} shows two examples of complex line profiles among our mock spectra. Our fitting results use fewer free parameters and have lower BIC values compared with the input parameters. In these circumstances, fewer free parameters than the input are statistically preferred, which means it is no longer possible to obtain the original parameters through any fitting procedure. Thus, there is little benefit in seeking the whole likelihood distribution in complex profiles (5-6 components), which is especially time-consuming using the MCMC method.
\end{enumerate}

The procedure of our Gaussian decomposition toolkit can be divided into two main parts: estimation of the initial guess (Section~\ref{subsubsection:Gaussian_deccomp_estimation}) and model selection using BIC (Section~\ref{subsubsection:BIC_selection}). The flow chart of the toolkit is shown in Fig.~\ref{fig:fitting_flow}. For a better illustration, an artificial spectrum was fitted as an example, the parameters of which are available in Table~\ref{tab:paramters_artificial_spec}.

\begin{table}
    \centering
    \caption{Parameters of the artificial spectrum in Figure.~\ref{fig:mock_example_spec}. The fitting results are shown in the output row. The colour of each component is labeled under the number.}
    \label{tab:paramters_artificial_spec}
    \begin{tabular}{llcccc}
        \hline
        Component & & 1 & 2 & 3 & 4 \\ 
        number & & (Cyan) & (blue) & (red) & (yellow) \\
        \hline
        velocity  & input  & 220  & 280  & 320 & 320         \\  
        (\kms\ ) & output & 220.24     & 279.92   & 319.86 & 340.36      \\ 
        sigma  & input  &10      & 15        & 15   & 100         \\ 
        (\kms\ ) & output & 11.66 & 15.34   & 15.04    & 103.54      \\ 
        amplitude  & input  & 0.6  & 1.6   & 0.9  & 2        \\ 
        (Jy \kms\ )  & output & 0.69 & 1.68 & 0.98   & 1.8        \\
        $\sigma _{rms}$ &   &                         & 1 mJy      & &    \\
        (input noise) &   &                         &  & &    \\
        velocity &    &       & 10 \kms\   &   &    \\
        resolution &    &       &    &   &    \\
        \hline
\end{tabular}
\end{table}

\subsubsection{Estimation of multiple Gaussian functions}
\label{subsubsection:Gaussian_deccomp_estimation}

To prevent the $\chi^{2}$ minimization from getting stuck in a local minimum, an estimation algorithm was developed to provide accurate initial guesses. The algorithm is based on a simple assumption: If a spectrum can be properly described by several Gaussian components, at least one component dominates some parts of the emission, and that component could be representative for those regions. For example, 3 components respectively dominate the 3 colour-filled regions in panel (1) of Fig.~\ref{fig:mock_example_spec}, and their parameters could be estimated if a Gaussian function were fitted to their representative regions (colour-filled regions). As a result, the whole emission could be decomposed in several iterations: in each iteration, one Gaussian component can be identified and subtracted if the representative regions of this component were found. 

The whole process can be divided into a preparation step and several iterations:

\begin{figure}
	\includegraphics[width=\columnwidth]{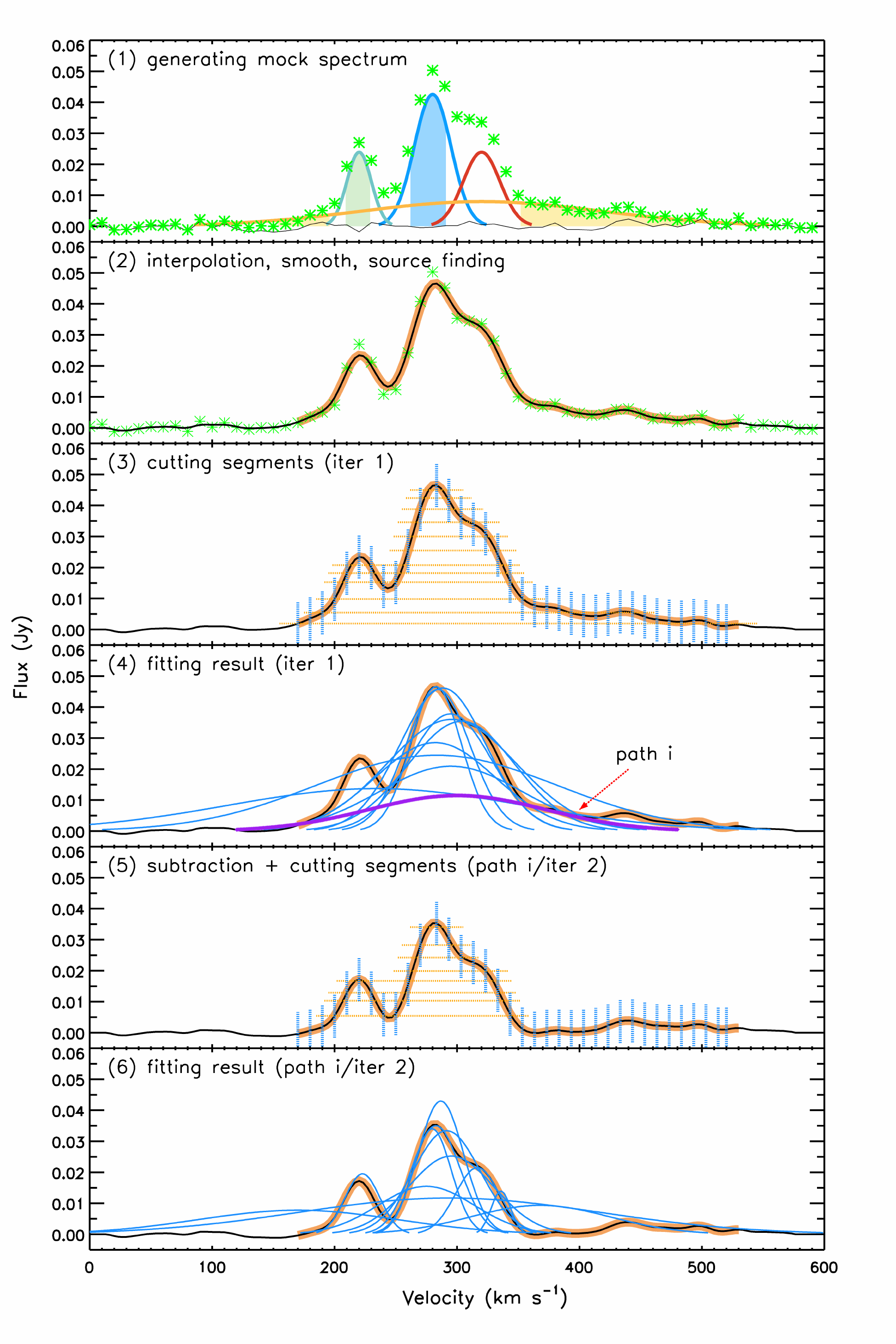}
    \caption{Example steps of initial guess estimation of the artificial spectrum. (1) The input Gaussian components (in coloured lines; the parameters of each component can be found in Table.~\ref{tab:paramters_artificial_spec} based on its color) and final spectrum (in green asterisk; with added Gaussian noise). The colour-filled areas refer to regions where at least 80\% of local emissions consist of a single component. Gaussian noise is plotted in the black line. There is no red area since component 3 (red line) overlaps with others and dominates no region. (2) The resampled spectrum (black line), source (orange thick line), and the original spectrum (green asterisk). (3) cutting segments results for iteration 1. Vertical lines refer to segments for 'base fit' and flat lines refer to segments for 'peak fit'. (4) All paths (in blue lines) generated by segment combinations from the last step. Path i is plotted in a thick purple line. (5) residual after path.i was subtracted. similarly, the residual.i is cut into segments. (6) all paths generated from segments combination at the last step.}
    \label{fig:mock_example_spec}
\end{figure}

\begin{enumerate}
\item Preparation: the input spectrum is interpolated (to a resolution of $\sim $ 3 \kms,\ which is sufficient for most extragalactic observations), smoothed (using a Gaussian PSF with a standard deviation of 6 \kms\ ), and masked (using a threshold of $\sim$ 2 times the rms) to generate an ideal spectrum for calculating initial guesses. The purpose of this step is to provide a new spectrum with better sampling and higher S/N value without strongly changing the shape of the spectrum. The source region (for segment cutting) was defined using a threshold of 2 $\sigma _{rms}$. Notably, the source region is only used for parameter estimation. All BIC values were calculated using data points in the original spectrum.
\item Iteration 1.1: the source region of the resampled spectrum is cut into a set of regularly spaced segments. There are two kinds of segments trying to resolve two different kinds of mixtures of Gaussian components. Peak segments refer to all resampled data points above their thresholds (shown as the horizontal dashed lines in panel (3) and panel (5) of Fig.~\ref{fig:mock_example_spec}). They are generated for situations where Gaussian components may be mixed with each other at the base while the shapes of their peaks remain recognizable (different central velocities), like component 2 (blue line) and 3 (red line) in panel (1) of Fig.~\ref{fig:mock_example_spec}. Base segments are the resampled data points within the small channel range between two adjacent vertical lines (such as the vertical dashed lines in panel (3) and panel (5) of Fig.~\ref{fig:mock_example_spec}). They are designed for situations where Gaussian components may be mixed with each other in the peak but part of their bases are still uncontaminated (same central velocity but different velocity dispersion), like component 3 (red) and 4 (yellow) in panel (1) of Fig.~\ref{fig:mock_example_spec}). (As shown in Table.~\ref{tab:paramters_artificial_spec}, component 3 and 4 have the same central velocity of 320 \kms\ but different velocity dispersion.) 
\item Iteration 1.2: Each peak segment or a combination of two base segments will be fitted by a Gaussian function. If the estimated Gaussian component (G.i) is not repetitive (it has the same Gaussian parameters as other estimations that have been already stored in this iteration) or exceeds the original spectrum (which will cause negative signals in the residual), an independent path will be generated, where G.i will be stored in. Then G.i will be subtracted from the input spectrum. The residual (Residual.i) will be stored and treated as the input of the path.i in iteration 2. For example, panel (4) of Fig.~\ref{fig:mock_example_spec} shows all G.i obtained by fitting to the segments shown in panel (3). Each G.i will yield a new Residual.i, which will be cut into segments and fitted in the next iteration. Notably, the total number of G.i in panel (4) is obviously smaller than the number of segment combinations in panel (3). This is because most of the segment combinations would not provide an acceptable solution. As a result, the number of paths is far less than the number of segment combinations.
\item Iteration 2: Each Residual.i obtained in iteration 1 will be treated independently as the input spectrum in this iteration. For instance, G.i (the thick purple line in panel (4) of Fig.~\ref{fig:mock_example_spec}) was subtracted, and its residual was cut and fitted in panel (5) and (6). Iterations will be repeated until the maximum number of Gaussian components (6 for this research) is reached or no reasonable path could be obtained from the residuals. 
\end{enumerate}
After several iterations, a few paths are found and stored. Since the number of Gaussian components stored in each path is different (those paths generated in iteration i will contain i components), paths containing the same number of Gaussian components are compared together using the observed spectrum. Since they are obtained by fitting to the segments of the resampled spectrum, they are good initial guesses, and the $\chi^{2}$ minimization fitting converges fast. Finally, the best solution for different numbers of components is stored. They will be compared using their BIC values, which will be described in the next section.

\subsubsection{Model selection using BIC}
\label{subsubsection:BIC_selection}
To select the best model for the spectrum, the BIC value is computed as follows:
\begin{equation}
BIC = k\cdot In(N) - 2* In(\hat{L}) \tag{3.1} 
\end{equation}

where k denotes the number of free parameters (3 for 1 Gaussian component), N refers to the number of data points, and $\hat{L}$ is the maximum value of the likelihood function. For Gaussian distributed noise, the likelihood function can be defined as follows:
\begin{equation} 
\label{eq_3.2}
L \propto \prod_{i=1}^{N} (exp( - (x_{i} -X_{i})^2 /2)) \tag{3.2}
\end{equation}

where $X_{i}$ is the observed value, while $x_{i}$ is the fitted value. Combining with equation~\ref{eq_3.2}, we then obtain:

\begin{equation}
BIC = k\cdot In(N) + \sum_{i=1}^{N} (x_{i} -X_{i})^2 \tag{3.3}
\end{equation}

Considering $\chi ^{2} = \sum_{i=1}^{N} (x_{i} -X_{i})^2 $, finally we get:

\begin{equation}
\label{eq_3.4}
BIC = k\cdot In(N) + \chi ^{2} \tag{3.4}
\end{equation}

\begin{figure}
	\includegraphics[width=\columnwidth]{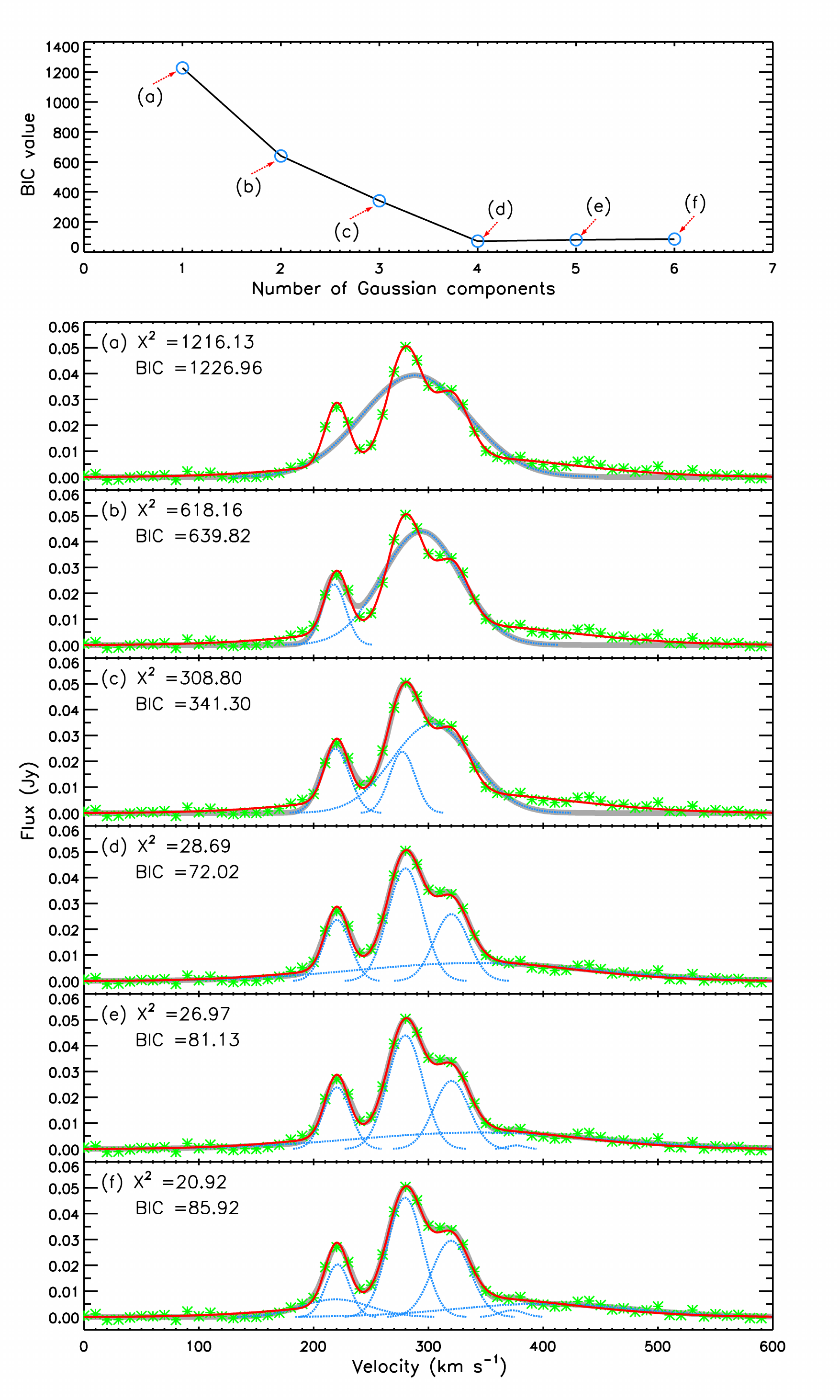}
    \caption{top panel: BIC value distribution of best solutions for different Gaussian components. bottom panels: The exact solution corresponding to each data point in the top panel. different Gaussian components (blue dashed lines), the best solution (thick grey lines), input spectrum (red lines), and observed data points with noise being added (green asterisk) were also plotted in each panel.}
    \label{fig:BIC_distribution}
\end{figure}

Under the assumption of Gaussian noise, equation~\ref{eq_3.4} connects the BIC value and $\chi ^{2}$ value. With more Gaussian components (free parameters) being allowed in a fitting procedure, the $\chi ^{2} $ value tends to decrease. Meanwhile, more free parameters will cause more BIC penalties as the first term of equation~\ref{eq_3.4} gets larger. We, therefore, accept the fit with the smallest BIC value as the final solution. For example, Fig.~\ref{fig:BIC_distribution} shows the BIC distribution of best solutions containing different numbers of Gaussian components, which were obtained in the previous section. With the number of Gaussian components increasing, the BIC value decreases to 72.02 at 4 Gaussian components. For more components, the BIC value rises again due to BIC penalties from more parameters although $\chi ^{2}$ keeps dropping, which suggests overfitting to the spectrum. By comparing the output (panel (d) of Fig.~\ref{fig:BIC_distribution}) with the input (panel (1) of Fig.~\ref{fig:mock_example_spec}), the fitting result successfully reproduced 4 input components. Details of the output parameters are available in Table~\ref{tab:paramters_artificial_spec}.

\bsp	
\label{lastpage}
\end{document}